\UseRawInputEncoding
\documentclass[journal, 10pt]{IEEEtran}
\usepackage{epstopdf}
\usepackage{mathrsfs}
\usepackage{amssymb}
\usepackage[mathcal]{euscript}
\usepackage{cite}
\usepackage{graphicx}
\usepackage{psfrag}
\usepackage{subfigure}
\usepackage{url}
\usepackage{stfloats}
\usepackage{amsmath}
\usepackage{array}
\usepackage{amsthm}
\usepackage{booktabs}  
\usepackage{algorithm} 
\usepackage{setspace}
\usepackage{algorithmic} 
\usepackage{amsmath,amsfonts,amssymb,amsbsy,bm,paralist,theorem,ifthen,color}
\usepackage{color}
\usepackage{graphicx}
\usepackage{color}
\usepackage{placeins}
\usepackage{float}
\usepackage{tabularx,colortbl}
\usepackage{multirow}
\usepackage{multicol}
\usepackage{arydshln}
\usepackage{makecell}
\begin{document}
\setlength{\textfloatsep}{4.5pt}
\title{Pervasive Wireless Channel Modeling Theory and Applications to 6G GBSMs for All Frequency Bands and All Scenarios}

\author{Cheng-Xiang Wang,~\IEEEmembership{Fellow,~IEEE}, Zhen Lv,~\IEEEmembership{Member,~IEEE}, Xiqi Gao,~\IEEEmembership{Fellow,~IEEE}, Xiaohu~You,~\IEEEmembership{Fellow,~IEEE}, Yang~Hao,~\IEEEmembership{Fellow,~IEEE}, and Harald~Haas,~\IEEEmembership{Fellow,~IEEE}

\thanks{Copyright (c) 2015 IEEE. Personal use of this material is permitted. However, permission to use this material for any other purposes must be obtained from the IEEE by sending a request to pubs-permissions@ieee.org.}

\thanks{This work was supported by the National Key R\&D Program of China under Grant 2018YFB1801101, the National Natural Science Foundation of China (NSFC) under Grant 61960206006, the High Level Innovation and Entrepreneurial Research Team Program in Jiangsu, the High Level Innovation and Entrepreneurial Talent Introduction Program in Jiangsu, and the EU H2020 RISE TESTBED2 project under Grant 872172.}

\thanks{C.-X. Wang (corresponding author), X. Q. Gao, and X.-H. You are with the National Mobile Communications Research Laboratory, School of Information Science and Engineering, Southeast University, Nanjing, 210096, China, and also with the Purple Mountain Laboratories, Nanjing, 211111, China  (email: $\left\{\text{chxwang, xqgao, xhyu}\right\}$@seu.edu.cn).}
\thanks{Z. Lv is with the Purple Mountain Laboratories, Nanjing, 211111, China  (email: lvzhen@pmlabs.com.cn).}
\thanks{Y. Hao is with the School of Electronic Engineering and Computer Science, Queen Mary University of London, London E1 4NS, U.K. (e-mail: y.hao@qmul.ac.uk).}
\thanks{H. Haas is with the LiFi Research and Development Center, Department Electronic and Electrical Engineering, The University of Strathclyde, Glasgow G1 1XQ, U.K. (e-mail: harald.haas@strath.ac.uk).}
\vspace{-0.6 cm}
}
\markboth{IEEE TRANSACTIONS ON VEHICULAR TECHNOLOGY, VOL. XX, NO. XX, MAY 2022}{}
\maketitle
\begin{abstract}
In this paper, a pervasive wireless channel modeling theory is first proposed, which uses a unified channel modeling method and a unified equation of channel impulse response (CIR), and can integrate important channel characteristics at different frequency bands and scenarios. Then, we apply the proposed theory to a three dimensional~(3D) space-time-frequency (STF) non-stationary geometry-based stochastic model~(GBSM) for the sixth generation~(6G) wireless communication systems. The proposed 6G pervasive channel model~(6GPCM) can characterize statistical properties of channels at all frequency bands from sub-6~GHz to visible light communication (VLC) bands and all scenarios such as unmanned aerial vehicle~(UAV), maritime, (ultra-)massive multiple-input multiple-output (MIMO), reconfigurable intelligent surface~(RIS), and industry Internet of things (IIoT) scenarios. By adjusting channel model parameters, the 6GPCM can be reduced to various simplified channel models for specific frequency bands and scenarios. Also, it includes standard fifth generation (5G) channel models as special cases. In addition, key statistical properties of the proposed 6GPCM are derived, simulated, and verified by various channel measurement results, which clearly demonstrates its accuracy, pervasiveness, and applicability.
\end{abstract}


\IEEEpeerreviewmaketitle

\section{Introduction}

With the commercialization of the fifth generation~(5G) wireless communication systems worldwide, many countries and organizations have started to conduct channel measurements and modeling for the sixth generation~(6G) wireless communication systems~\cite{6G_SCI}. Comprehensive surveys have summarized different types of 6G wireless channels by grouping them under all spectra, global-coverage scenarios, and full-application scenarios~\cite{6G_VTM, 6G_SCI}. Consequently, 6G wireless channels exhibit many new channel characteristics and entail new requirements for 6G channel modeling.

In the 6G wireless communication systems, all spectra can be utilized, including sub-6 GHz, millimeter wave~(mmWave), terahertz~(THz), and optical wireless frequency bands~\cite{HeRS_mmWave_TVT18,GuanK_THz_TVT21,Zhu_VLC}. MmWave and THz channels show new characteristics, e.g., large bandwidth with high delay resolution, frequency non-stationarity, high directivity, diffuse scattering, blockage effects, and atmosphere absorption~\cite{HuangJ2020_JSAC, Liu2019_HST, THz_VTM2021, THz_WangJ}. Visible light communication (VLC) channels have no small-scale fading~(SSF), negligible Doppler effect, and frequency non-stationarity~\cite{Zhu_VLC}. To achieve global coverage in the 6G wireless communication systems, in addition to terrestrial communications, satellite, unmanned aerial vehicle~(UAV), and maritime communications should also be considered~\cite{LEO_measurement, UAV_TVT2021, Maritime_JSAC2021}. In low Earth orbit~(LEO) satellite communication channels, the most remarkable channel characteristics are large Doppler shift caused by the rapid movement of satellites~\cite{LEO_measurement}, the influence of meteorological factors such as rain attenuation~\cite{LEO_waterVapor}, and ionosphere effects, e.g., Faraday rotation~\cite{LEO_faradayrotation}. Furthermore, in UAV channels, arbitrary three dimensional~(3D) trajectories of UAV and altitudes-dependent large-scale parameters~(LSPs) should be considered~\cite{Chang2020_IoT, YinXF2021_UAV}. In maritime communication channels, we need to take into account the fluctuation of sea waves and the location-dependent property, i.e., components generated by sea wave scattering, evaporation duct propagation, and line-of-sight~(LoS) propagation will appear or disappear according to the distance between the transmitter (Tx) and the receiver (Rx)~\cite{HeY_Maritime}. In terms of full-application scenarios, vehicle-to-vehicle~(V2V), high-speed train~(HST), vacuum tube ultra-high-speed train~(UHST), (ultra-)massive multiple-input multiple-output~(MIMO), reconfigurable intelligent surface~(RIS), and industry Internet of things~(IIoT) communications stand a good chance of being utilized in the 6G wireless communication systems. V2V channels are featured by large Doppler shift, temporal non-stationarity caused by fast-changing environments, and multiple mobilities of Tx, Rx, and scatterers~\cite{BianJ_V2V19}. When it comes to ultra-high mobility scenarios such as UHST scenarios, the underlying channels experience much larger Doppler shift and stronger temporal non-stationarity~\cite{Zhang2019_HST, GuanK2018_HST}. The waveguide effect and the impact of tube wall roughness should be taken into account in vacuum tube UHST scenarios~\cite{Xu2021_UHST}. The spherical wavefront and non-stationarity in the spatial domain need to be considered in (ultra-)massive MIMO channels. RIS-based 6G wireless channels need to consider cascaded sub-channels separated by a RIS and phase shift matrix according different RIS designs. Besides, channels in IIoT scenarios show characteristics such as rich scattering and multi-mobility property caused by the existence of a large number of moving mechanical equipment~\cite{6G_VTM,LiY_PIMRC}.

The challenge of channel modeling for the 6G wireless communication systems is how to combine those channel characteristics into a unified framework, since there are usually mixed applications of various new technologies. For example, multi-link mmWave wireless communication systems may use ultra-massive MIMO under high-mobility scenarios such as V2V, (U)HST, and UAV scenarios. This communication channel will show significant space-time-frequency~(STF) non-stationarity~\cite{3D_nostationary}. Meanwhile, spatial consistency~\cite{3GPP38.901} should be considered in most channel models. Consequently, constructing realistic channel models with the best trade-off among accuracy, complexity, and pervasiveness is urgently and essentially needed in the beginning of 6G research.

To the best of our knowledge, there is no existing model taking all aforementioned properties into account. B5GCM~\cite{Bian2021_B5G} and several standard 5G channel models~\cite{comst_5GSurvey},~i.e., 3GPP TR~38.901~\cite{3GPP38.901}, IMT-2020~\cite{IMT2020}, and QuaDRiGa~\cite{QuaDRiGa} have made some efforts. However, for all-spectra cases, they all neglected VLC band and more or less ignored part of the characteristics at mmWave and THz bands. For example, QuaDRiGa channel model ignored gas absorption and blockage effect, 3GPP TR~38.901 and IMT-2020 channel models did not consider frequency non-stationarity. For global-coverage scenarios, the above-mentioned channel models were only suitable for terrestrial wireless communication scenarios without considering satellite, UAV, and maritime communication scenarios. For full-application scenarios, exiting channel models mentioned above could not support vacuum tube UHST and RIS scenarios. Besides, 3GPP TR~38.901 and IMT-2020 channel models omitted spherical wavefront and spatial non-stationarity in (ultra-)massive MIMO scenarios. It follows that B5GCM and standard 5G channel models still lack pervasiveness.

To fill the research gap, this paper aims to propose a pervasive wireless channel modeling theory and construct a 6G pervasive channel model~(6GPCM) by applying this theory to a geometry-based stochastic model (GBSM). The novelties and main contributions of this paper are summarized as follows:
\begin{enumerate}
  \item A pervasive wireless channel modeling theory is first proposed, which uses a unified channel modeling method and a unified equation of channel impulse response (CIR) to model all important channel characteristics in all frequency bands and all scenarios including global-coverage and full-application scenarios.
  \item The proposed 6GPCM is the direct application of the pervasive wireless channel modeling theory to 6G with a GBSM framework. It considers channel characteristics in all spectra from sub-6 GHz to VLC bands, global-coverage scenarios including LEO satellite, UAV, and maritime communication channels, and full-application scenarios such as (U)HST, (ultra-)massive MIMO, RIS, and IIoT communication channels in the 6G wireless communication systems.
  \item The proposed 6GPCM can easily be simplified to a specific channel model for a specific frequency band and/or a specific scenario by adjusting channel model parameters. Therefore, the 6GPCM includes many existing channel models, such as standard 5G channel models, as special~cases.
  \item Key statistical properties of the 6GPCM have been derived, simulated, and compared with many channel measurements at specific frequency bands and scenarios, showing the correctness of derivation and simulation results, accuracy, pervasiveness, and applicability of the proposed 6GPCM.
\end{enumerate}

The remainder of this paper is organized as follows. In Section~\ref{Sec_2} and Section~\ref{Sec_3}, the pervasive wireless channel modeling theory and the 6GPCM are illustrated, respectively. Statistical properties of the 6GPCM are derived and analyzed in Section~\ref{Sec_4}. Analytical, simulation, and some measurement results are illustrated and compared in Section~\ref{Sec_5}. Finally, conclusions are drawn in Section~\ref{sec_Conclusions}.

\section{Pervasive Wireless Channel Modeling Theory}
\label{Sec_2}
\begin{figure}[t]
    \setlength{\abovecaptionskip}{0.1cm}
	\centering\includegraphics[width=0.49\textwidth]{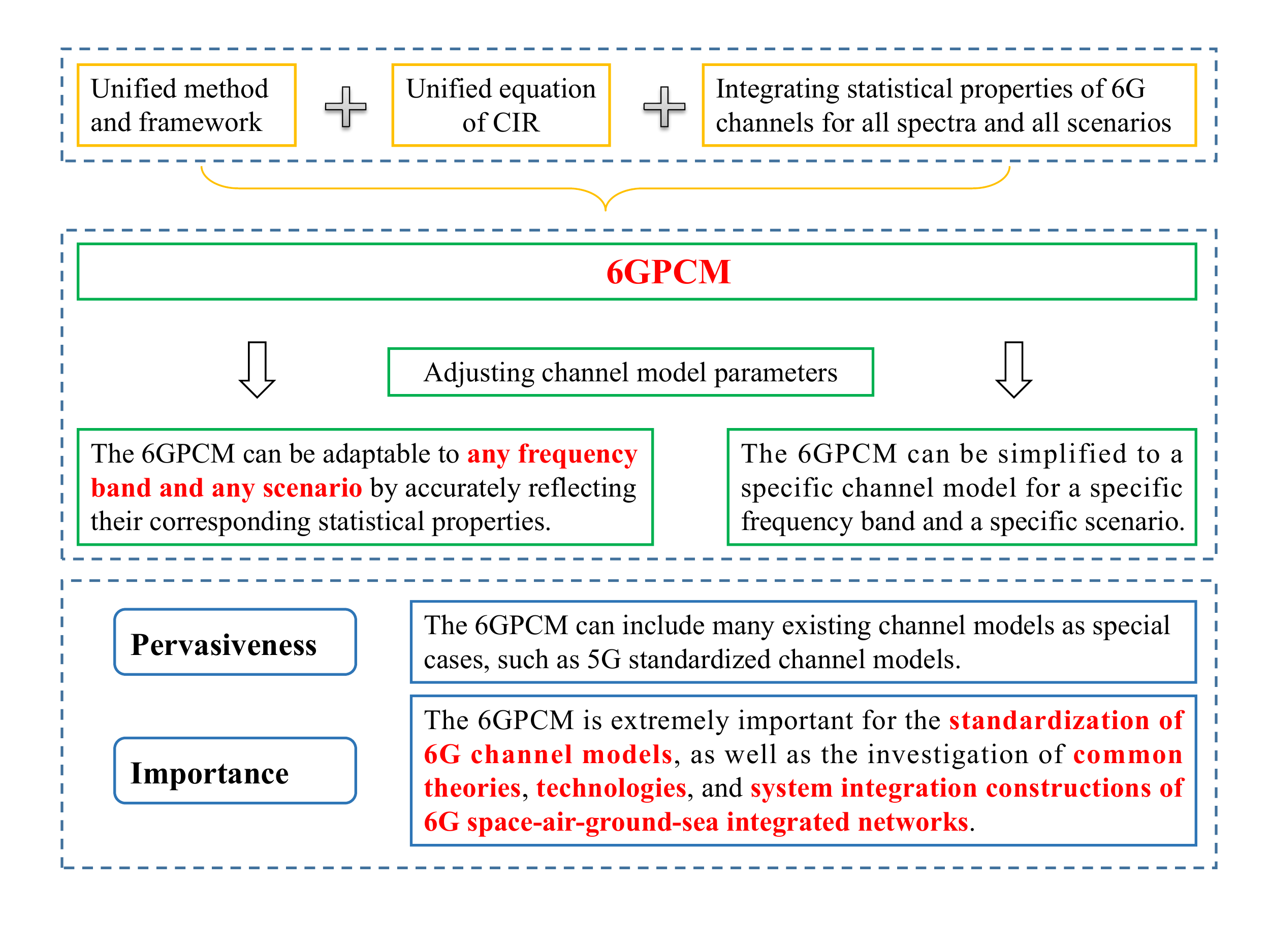}
	\centering\caption{The 6GPCM for 6G all frequency bands and all scenarios.}
	\label{fig_pervasive}
\end{figure}

The pervasive wireless channel modeling theory aims to use  a unified channel modeling framework and a unified equation of CIR to characterize channel properties of all frequency bands and all scenarios, as shown in Fig.~\ref{fig_pervasive}. Meanwhile, the pervasive wireless channel model should also be adaptable to specific channels at any frequency band and any scenario by adjusting channel model parameters. Therefore, the pervasive channel model includes many existing channel models as special cases. Applying this theory to a 6G GBSM, we can obtain a geometry-based stochastic 6GPCM. Similarly, we can propose a deterministic 6GPCM, which considers 6G wireless channel characteristics using a ray tracing method. We can also propose a predictive 6GPCM based on a machine learning method~\cite{HuangC_TVT2020} utilizing this theory.

With a 6GPCM, the complex mapping relationships among 6G channel model parameters, channel characteristics, and communication system performance can be studied. Also, different technologies and antenna arrays can be compared since 6GPCM has a unified framework supporting various new technologies and any antenna arrays with different sizes, structures, and radiation patterns. Consequently, it is extremely important for the standardization of 6G channel models, researches on common theories and technologies, and system integration constructions of 6G space-air-ground-sea integrated networks.

In this paper, we propose a 6GPCM, which is a 3D GBSM, based on the pervasive wireless channel modeling theory. It considers channel characteristics for all spectra, global-coverage scenarios, and full-application scenarios for the 6G wireless communication systems. Also, the proposed 6GPCM can be simplified to specific channel models at any specific frequency band and/or any specific scenario by adjusting channel model parameters. The details are as follows:
\begin{enumerate}
  \item Geometry-based stochastic modeling method and framework: Firstly, we generate spatial correlated LSPs. Then, an ellipsoid Gaussian scattering model~\cite{Bian2021_B5G} is introduced to model position coordinates of rays in each cluster according to values of LSPs. After that, no matter how the locations of Tx, Rx, and clusters are moved, we can obtain the values of delays and angles of rays in the clusters at each snapshot according to geographical locations of the Tx, Rx, and clusters. When delays are known, the powers can be calculated correspondingly.
  \item The unified equation of CIR:  We use a unified equation of CIR to characterize channel properties of all frequency bands and all scenarios.
  \item Integrating statistical properties of 6G channels: We consider channel characteristics in all spectra, i.e., $\text{sub-6}$ GHz, mmWave, THz, and VLC bands, global-coverage scenarios including LEO satellite, UAV, and maritime communication scenarios, and full-application scenarios such as (U)HST, (ultra-)massive MIMO, RIS, and IIoT scenarios in the 6G wireless communication systems. The details are explained in Section~\ref{Sec_3}.
\end{enumerate}

\section{The 6GPCM Based on the Pervasive Channel Modeling Theory}
\label{Sec_3}
\begin{figure}
\setlength{\abovecaptionskip}{-0.4cm}
\centering\includegraphics[width=0.9\columnwidth]{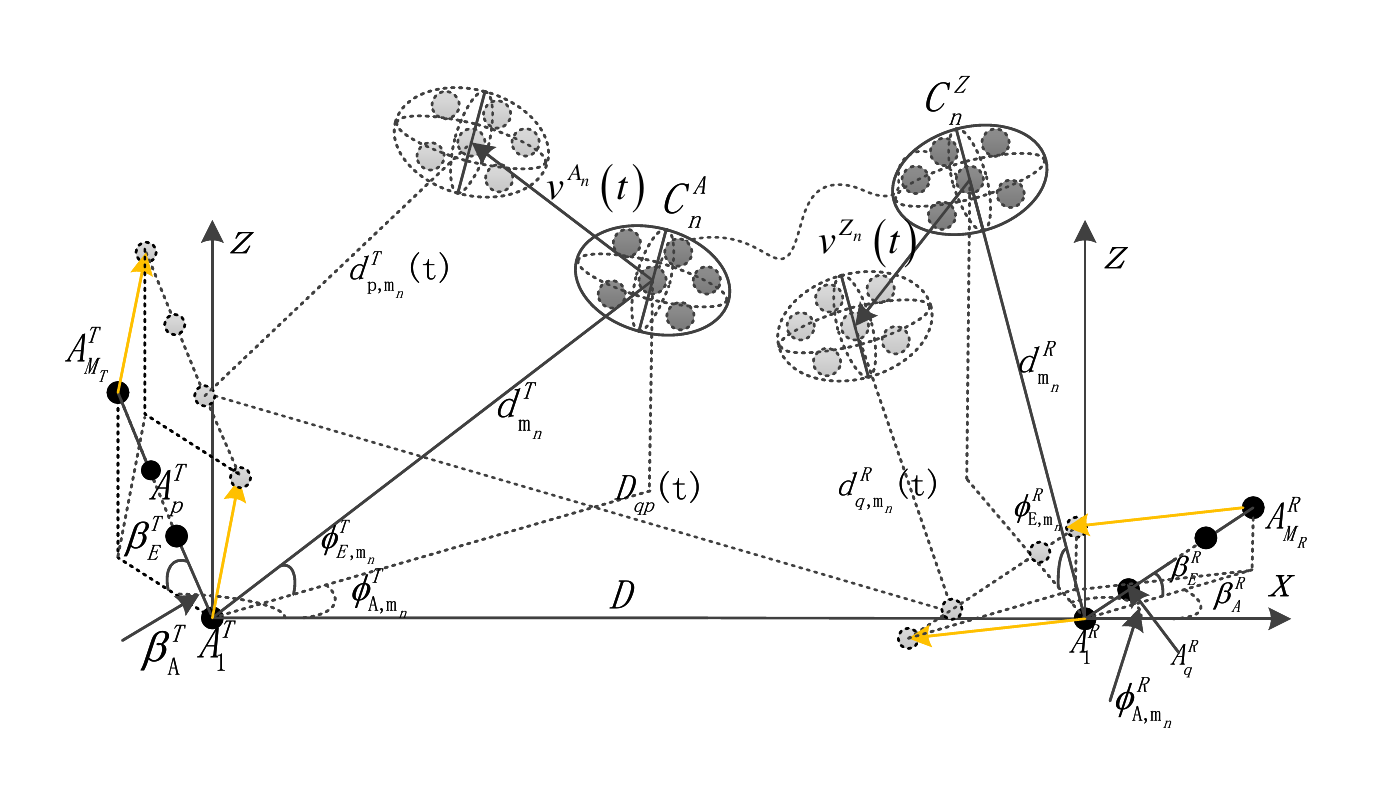}
\caption{Illustration of the 6GPCM.}
\label{6gchannels}
\vspace{0.3cm}
\end{figure}

The 6GPCM, which is illustrated in Fig.~\ref{6gchannels}, is a pervasive MIMO channel model with multiple Tx-Rx links at multiple frequencies,~i.e., there are $N_T$ ($N_R$) Tx (Rx) antenna arrays at multiple carrier frequencies. Note that our model supports both uniform planar arrays~(UPAs) and uniform linear arrays~(ULAs). ULAs are employed at both the Tx and the Rx sides in Fig.~\ref{6gchannels}, where each Tx (Rx) is equipped with $M_T$ ($M_R$) antenna elements, $A_p^T$ ($A_q^R$) means the $p$th ($q$th) antenna element spaced at $\delta_T$ ($\delta_R$), $\beta_A^{T(R)}$ is the azimuth angle of Tx (Rx) antenna array, whereas $\beta_E^{T(R)}$ is the elevation angle of the Tx (Rx) antenna array. In order to make the illustrations clear, only the $n$th ($n=1,...,N_{qp}(t)$) cluster pair is shown when considering multi-bounce propagation in this figure, where $C_n^A$ is the first-bounce cluster of the $n$th cluster pair at the Tx side and $C^Z_n$ is the last-bounce cluster of the $n$th cluster pair at the Rx side. Note that unless otherwise stated, ``cluster" in this paper refers to ``cluster pair" in a multi-bounce channel model and ``cluster" in a single-bounce channel model. The propagation between $C^A_n$ and $C^Z_n$ is abstracted by a virtual link with a random delay~\cite{WINNERII}. When the delay of the virtual link is set to zero, the cluster pair will completely overlap with each other and the multi-bounce channel model will be reduced to a single-bounce channel model~\cite{Bian2021_B5G}. Besides, $N_{qp}(t)$ is the number of clusters from $A^T_p$ to $A^R_q$, and $M_n(t)$ is the number of rays following Poisson distributions in the $n$th cluster at time instant $t$~\cite{Liu2019_HST}. The 6GPCM supports 3D arbitrary trajectory and multi-mobility property, the movements of Tx, Rx, and clusters are described by the speed $v^X(t)$, travel azimuth angles $\alpha_A^{X}(t)$, and travel elevation angles $\alpha_E^{X}(t)$. The superscript $X\in\{T,R,A_n,Z_n\}$ denotes the Tx, Rx, $C_n^A$, and $C_n^Z$, respectively. For clarity, key parameters are listed in Table~\ref{par}.
\subsection{CIR}
The complete channel matrix of the 6GPCM can be expressed as
  \begin{align}
  \label{CIR_singleLink}
  \mathbf{H}=\left [PL \cdot SH \cdot BL \cdot WE \cdot AL\right ]^{1/2}\cdot \mathbf{H_s}
  \end{align}
where $PL$ denotes the path loss caused by the propagation distance between Tx and Rx. Shadowing is represented by $SH$, $BL$ denotes the blockage loss~\cite{mmWave_HuangJ_JSAC17}, $WE$ means the weather effect loss, e.g., rain attenuation loss~\cite{LEO_RA} in LEO satellite communication scenarios, and $AL$ represents the atmospheric gas absorption loss, such as the oxygen absorption loss at mmWave band~\cite{IMT2020} and the molecular absorption loss at THz band~\cite{Barros2017_THz_OL,WangJ_THz_TVT21}. All these large-scale fading~(LSF) parameters discussed are calculated in power level.

The SSF can be represented by the matrix $\mathbf{H_s}=\left[h_{q p, f_{c}}(t, \tau)\right]_{M_{R} \times M_{T}}$, where $h_{q p, f_{c}}(t, \tau)$ is the CIR between $A_p^T$ and $A_q^R$ at the carrier frequency $f_c$. Specifically, $h_{q p, f_{c}}(t, \tau)$ can be represented by the superposition of the LoS and non-LoS~(NLoS) components,~i.e.,
\begin{table*}[t!]\footnotesize
   \caption{Definitions of Significant Parameters.}
   \center
   \label{par}
   \begin{spacing}{1.1}
    \begin{tabular}{|c|p{14cm}|}
      \hline
      \text{Parameters} & \makecell[c]{\text{Definitions}}\\
      \hline
      $\vec{A}_p^T(t)/\vec{A}_q^R(t)$   &  Coordinate of $p$/$q$th Tx/Rx antenna element in global coordinate at time instant $t$\\ \hline
      $\vec{C}^A_n/\vec{C}^Z_n$   &  Coordinate of center of $C^A_n/C^Z_n$ relative to Tx/Rx\\ \hline
      $\vec{C}^A_{m_n}/\vec{C}^Z_{m_n}$   &  Coordinate of the $m$th scatterer in $C^A_n/C^Z_n$\\ \hline
      $\beta^{T}_A/\beta^{T}_E/\delta_T$   &   Azimuth angle, elevation angle, and antenna spacing of the Tx antenna array \\ \hline
      $\beta^{R}_A/\beta^{R}_E/\delta_R$   &   Azimuth angle, elevation angle, and antenna spacing of the Rx antenna array \\ \hline
      $\phi^T_{A,m_n}(t)$ & Azimuth angle of departure (AAoD) of $C^A_{m_n}$ at time instant $t$\\ \hline
      $\phi^T_{E,m_n}(t)$ & Elevation angle of departure (EAoD) of $C^A_{m_n}$ at time instant $t$\\ \hline
      $\phi^R_{A,m_n}(t)$ & Azimuth angle of arrival (AAoA) of $C^Z_{m_n}$ at time instant $t$\\ \hline
      $\phi^R_{E,m_n}(t)$ & Elevation angle of arrival (EAoA) of $C^Z_{m_n}$ at time instant $t$\\ \hline
      $\phi^T_{A,\mathrm{LoS}}(t)/\phi^T_{E,\mathrm{LoS}}(t)$ & AAoD and EAoD of the LoS component at time instant $t$\\ \hline
      $\phi^R_{A,\mathrm{LoS}}(t)/\phi^R_{E,\mathrm{LoS}}(t)$ & AAoA and EAoA of the LoS component at time instant $t$\\ \hline
      $\alpha^T_A(t)/\alpha^T_E(t)$ & Travel azimuth/elevation angles of the Tx antenna array at time instant $t$\\ \hline
      $\alpha^R_A(t)/\alpha^R_E(t)$ & Travel azimuth/elevation angles of the Rx antenna array at time instant $t$\\ \hline
      $\alpha^{A_n}_A(t)/\alpha^{A_n}_E(t)$ & Travel azimuth/elevation angles of $C^A_n$ at time instant $t$\\ \hline
      $\alpha^{Z_n}_A(t)/\alpha^{Z_n}_E(t)$ & Travel azimuth/elevation angles of $C^Z_n$ at time instant $t$\\ \hline
      $v^T(t)/v^R(t)$ & Speed value of the Tx/Rx antenna array at time instant $t$\\ \hline
      $v^{A_n}(t)/v^{Z_n}(t)$ & Speed value of $C^A_n$/$C^Z_n$ at time instant $t$\\ \hline
      $D$/$D_{qp}(t)$ & Distance between  $A_p^T$ and $A_q^R$ at initial time/time instant $t$ \\ \hline
    \end{tabular}
    \end{spacing}
    \vspace{-0.3cm}
\end{table*}
\small
\begin{align}
h_{q p, f_{c}}(t, \tau)=\sqrt{\frac{K_{R}(t)}{K_{R}(t)+1}} h_{q p, f_{c}}^\text{LoS}(t, \tau)+\sqrt{\frac{1}{K_{R}(t)+1}} h_{q p, f_{c}}^\text{NLoS}(t, \tau)
\label{CIR}
\end{align}
\normalsize
where $K_{R}(t)$ is the K-factor at time instant $t$, the calculations of $h_{q p, f_{c}}^\text{LoS}(t, \tau)$ and $h_{q p, f_{c}}^\text{NLoS}(t, \tau)$ can be expressed as~(\ref{CIR_LOS}) and~(\ref{CIR_NLOS}), respectively.
\begin{figure*}[b!]
\small
\hrule
\begin{align}
\label{CIR_LOS}
h_{q p, f_{c}}^\text{LoS}(t, \tau)= \left[\begin{array}{c}
{F_{q, f_{c}, V}\left(\phi_{E, \mathrm{L}}^{R}(t), \phi_{A, \mathrm{L}}^{R}(t)\right)}\\
{F_{q, f_{c}, H}\left(\phi_{E, \mathrm{L}}^{R}(t), \phi_{A, \mathrm{L}}^{R}(t)\right)}
\end{array}\right]^{{T}}
\left[\begin{array}{cc}
{e^{j \theta_{\mathrm{L}}^{V V}}} & 0 \\
0 & {e^{j \theta_{\mathrm{L}}^{HH}}}
\end{array}\right]\mathbf{F}_{\mathrm{r}}
\left[\begin{array}{cc}
{F_{p, f_{c}, V}\left(\phi_{E, \mathrm{L}}^{T}(t), \phi_{A, \mathrm{L}}^{T}(t)\right)}\\
{F_{p, f_{c}, H}\left(\phi_{E, \mathrm{L}}^{T}(t), \phi_{A, \mathrm{L}}^{T}(t)\right)}
\end{array}\right]
\cdot e^{j 2 \pi f_{c} \tau_{q p}^{L}(t)}  \delta\left(\tau-\tau_{q p}^{L}(t)\right)
\end{align}
\begin{align}
\label{CIR_NLOS}
\nonumber h_{q p, f_{c}}^\text{NLoS}(t, \tau) & =\sum_{n=1}^{N_{q p}(t)} \sum_{m=1}^{M_{n}(t)}\left[\begin{array}{c}
{F_{q, f_{c}, V}\left(\phi_{E, m_{n}}^{R}(t), \phi_{A, m_{n}}^{R}(t)\right)}\\
{F_{q, f_{c}, H}\left(\phi_{E, m_{n}}^{R}(t), \phi_{A, m_{n}}^{R}(t)\right)}
\end{array}\right]^{{T}}\left[\begin{array}{cc}
{e^{j \theta_{m n}^{V V}}}{ \, \, \,  \, \,\,\sqrt{\mu \kappa_{m_{n}}^{-1}(t)} e^{j \theta_{m_{n}}^{VH}}} \\
{ \sqrt{\kappa_{m_{n}}^{-1}(t)} e^{j \theta_{m_{n}}^{HV}}} \, \, \, \, \, \, \sqrt{\mu} {e^{j \theta_{m_{n}}^{HH}}}
\end{array}\right]
\mathbf{F}_{\mathrm{r}}\left[\begin{array}{cc}
{F_{p, f_{c}, V}\left(\phi_{E, m_{n}}^{T}(t), \phi_{A, m_{n}}^{T}(t)\right)}\\
{F_{p, f_{c}, H}\left(\phi_{E, m_{n}}^{T}(t), \phi_{A, m_{n}}^{T}(t)\right)}
\end{array}\right] \\
& \sqrt{P_{q p, m_{n}, f_{c}}(t)} \cdot e^{j 2 \pi f_{c} \tau_{q p, m_{n}}(t)} \cdot \delta\left(\tau-\tau_{q p, m_{n}}(t)\right)
\end{align}
\normalsize
\vspace{-0.3 cm}
\end{figure*}
Here, $\left\{\cdot\right\}^T$ stands for transposition operation, $F_{p(q), f_{c}, V}$ and $F_{p(q), f_{c}, H}$ are the antenna patterns of Tx (Rx) antenna for vertical and horizontal polarizations at corresponding carrier frequency $f_c$, respectively. Our model is independent of radiation patterns of antenna elements, i.e., any radiation pattern can be used here, $\kappa_{m_n}(t)$ is the cross polarization power ratio, $\mu$ is co-polar imbalance~\cite{Bian2021_B5G}, $\theta^{VV}_{m_n}$, $\theta^{VH}_{m_n}$, $\theta^{HV}_{m_n}$, and $\theta^{HH}_{m_n}$ are initial phases modeled as random variables uniformly distributed over $(0,2\pi]$. In addition, $\mathbf{F}_{\mathrm{r}}=\begin{bmatrix}    \text{cos}(\psi_{m_n}) & -\text{sin}(\psi_{m_n})\\  \text{sin}(\psi_{m_n}) & \text{cos}(\psi_{m_n})    \end{bmatrix}$ represents Faraday rotation referring to the rotation of the polarization plane caused by the propagation of electromagnetic waves through the ionosphere in LEO satellite scenario, and $\psi_{m_n} = 108 / {f^2_{c}}$ is the Faraday rotation angle~\cite{3GPP38.811}, where $f_c$ is in GHz. Otherwise, in scenarios without considering the influence of ionosphere, we can set $\psi_{m_n} = 0$. Additionally, $P_{q p, m_{n},f_{c}}(t)$ and $\tau_{q p, m_{n}}(t)$ is the power and delay of the $m$th ray in the $n$th cluster between $A_p^T$ and $A_q^R$ at time instant $t$. Also, $\theta_{\mathrm{L}}^{V V}$ and $\theta_{\mathrm{L}}^{H H}$ denote random phase in $(0,2\pi]$, $\tau_{q p}^{L}(t)$ is the time delay of LoS path at time instant $t$, and given the speed of light $c$, $\tau_{q p}^{L}(t)$ can be calculated as $\tau_{qp}(t)=\left \|\vec{A}^{R}_{q}(t) - \vec{A}^{T}_{p}(t)\right \|/c = D_{qp}(t)/c$, where $\left \| \cdot \right \|$ calculates the Frobenius norm.
Note that the delays are unresolvable under small bandwidth condition, we no longer model rays within clusters. We can replace $\tau_{q p, m_{n}}(t)$ and $P_{q p, m_{n}, f_{c}}(t)$ in~(\ref{CIR_NLOS}) with cluster delay $\tau_{q p, n}(t)$ and cluster power $P_{q p, n, f_{c}}(t)$.

When it comes to maritime ship-to-ship communication scenarios, the LoS path component and multipath components of both rough ocean surface and evaporation duct over the sea surface cannot be ignored. In our model, these three sub-parts in the maritime channel are modeled as $h_{q p, f_{c}}^\text{LoS}(t, \tau)$, $h_{q p, f_{c}}^{\text{NLoS}_1}(t, \tau)$, and $h_{q p, f_{c}}^{\text{NLoS}_2}(t, \tau)$. Then we use the power control factors $S_1$ and $S_2$ to manipulate the disappearance and appearance of the corresponding sub-parts with the change of the distance between the two ships, where $S_1 + S_2 = 1$. In IIoT scenarios, apart from the LoS component, specular multipath components (SMCs) and dense multipath components~(DMCs) are modeled as $h_{q p, f_{c}}^{\text{NLoS}_\text{SMC}}(t, \tau)$ and $h_{q p, f_{c}}^{\text{NLoS}_\text{DMC}}(t, \tau)$, respectively. The accuracy of channel models can be significantly improved by considering DMCs, which are caused by smaller scatterers and these scatterers are distributed near SMCs~\cite{LiY_PIMRC}. Although $h_{q p, f_{c}}^{\text{NLoS}_1}(t, \tau)$, $h_{q p, f_{c}}^{\text{NLoS}_2}(t, \tau)$, $h_{q p, f_{c}}^{\text{NLoS}_\text{SMC}}(t, \tau)$, and $h_{q p, f_{c}}^{\text{NLoS}_\text{DMC}}(t, \tau)$ have different forms, computing methods of them are the same as that of $h_{q p, f_{c}}^\text{NLoS}(t, \tau)$ with the same framework, different parameter values, and different distribution ranges of scatterers.

Besides, different from traditional radio frequency (RF) communication systems, the transmitted optical signals in VLC systems have extremely short wavelength~\cite{BL_VLC} and no phase since light-emitting diode (LED) lights emit incoherent light~\cite{Zhu_VLC}. It follows that the superposition of real-valued multipath signals at the Rx will not lead to SSF, but show slow-changing shadowing. Therefore, the 6GPCM in VLC bands is essentially a large-scale model, i.e., $\mathbf{H_s} = 1$. Correspondingly, we focus on modeling the path loss and shadowing in~(\ref{CIR_singleLink}) for VLC bands, as $\mathbf{H}= PL \cdot SH = \left[h_{q,p_{H}p_{V},\lambda_T}(t, \tau)\right]_{N_\text{PD}\times M_{p_H} \times M_{p_V}}$, where Tx is a $M_{p_H} \times M_{p_V}$ uniform planar LED array and Rx are $N_\text{PD}$ photodiodes (PDs). Here, $h_{q,p_Hp_V,\lambda_T}(t, \tau)$ is the CIR of the sub-channel between the $q$th PD and the LED element at the $p_H$th row and the $p_V$th column. Also, $\lambda_T$ represents the wavelength range of the light source. Details can be referred to~\cite{Zhu_VLC}. Note the fact that signals transmitted in VLC channels are real-valued optical power signals instead of complex amplitude signals in RF-based channels.

In multi-link channels, we assume $\mathbf{H_M}$ denotes the complete channel matrix, which can be obtained as
\begin{align}
\label{CIR_multiLink}
\mathbf{H_M} = {\left[ {\begin{array}{*{20}{c}}
{\mathbf{H}_{{1,1}}}& \cdots &{\mathbf{H}_{{1,{N_R}}}}\\
 \vdots & \ddots & \vdots \\
{\mathbf{H}_{{{N_T},1}}}& \cdots &{\mathbf{H}_{{{N_T},{N_R}}}}
\end{array}} \right]_{{N_T} \times {N_R}}}
\end{align}
where $\mathbf{H}_{i,j}$ is the channel matrix in the single-link channel, $i = 1, 2, \cdots, N_T$, and $j = 1, 2, \cdots, N_R$. The 6GPCM takes into consideration the spatial consistency~\cite{3GPP38.901}, including the spatial consistency of LSPs and small-scale parameters (SSPs). After considering spatial correlated parameters, single-link channel models can be extended to multi-link channel models~\cite{Zhang2020_WCSP_MF}.

\subsection{Generation of Spatially Correlated LSPs}
\label{subsec:LSPs}
LSPs include delay spread, K-factor, $SH$, elevation spread of departure~(ESD), elevation spread of arrival~(ESA), azimuth spread of departure~(ASD), azimuth spread of arrival~(ASA), and cross-polarization ratio~(XPR)~\cite{QuaDRiGa}. Taking generation of the delay spread for instance, it is firstly generated according to Gaussian process with corresponding mean value $ \text{DS}_{{\mu ,{f_c}}} $ and standard deviation $ \text{DS}_{{\sigma ,{f_c}}} $ at the frequency $ f_c $~(GHz), as
\begin{align}
\text{DS}_{{{f_c}}}\left( \mathbf{P} \right) = \text{DS}_{{\mu ,{f_c}}} + {X^{\text{DS}}}\left( \mathbf{P} \right) \cdot \text{DS}_{{\sigma ,{f_c}}}.
\end{align}
Here, $\mathbf{P} = (x_T, y_T, z_T, x_R, y_R, z_R)$ is the position of the Tx and the Rx, and $\text{DS}_{{{f_c}}}\left( \mathbf{P} \right)$ denotes the delay spread value at frequency $f_c$ and the position $\mathbf{P}$. Using the sum of sinusoids~(SoS) method in~\cite{QuaDRiGa}, we can get spatially correlated variable ${X^{\text{DS}}}\left( \mathbf{P} \right)\sim N(0,1)$, which ensures the continuity of parameter variation within correlation distance $d_{\text{corr\_DS}}$. Then, reference values of $ \text{DS}_{{\mu ,{f_c}}} $  and $ \text{DS}_{{\sigma ,{f_c}}} $ can be divided into three different configuration values according to the height of user terminal (UT) $h_\text{UT}$. For example, for terrestrial channel models $\left(1.5\,{\rm{m}} \le {h_\text{UT}} \le 22.5\,{\rm{m}}\right)$ in urban macro~(UMa) scenario, the parameters set from Table~7.5-6 of 3GPP TR~38.901~\cite{3GPP38.901} are recommended. For UAV channel model $\left(22.5\,{\rm{m}} \le {h_\text{UT}} \le 300\,{\rm{m}}\right)$ in UMa scenario, where $h_\text{UT}$ is the UAV altitude $h_\text{UAV}$, we prefer to get the parameters set of the corresponding scenario from Table B.1.2 of 3GPP TR~ 36.777~\cite{3GPP36.777} . In addition, we can turn to Table 6.7.2 of 3GPP TR~38.811~\cite{3GPP38.811} to get the corresponding parameters set for LEO satellite channel model in urban scenario. Taking the value of $ \text{DS}_{{\mu ,{f_c}}} $ at NLoS condition in UMa scenario and $f_c$ at S band~(2$\sim$4~GHz) for example, $ \text{DS}_{{\mu ,{f_c}}} $ is defined as
\begin{equation}
\text{lgDS}_{\mu ,{f_c}}\!=\!\left\{ {\begin{array}{*{18}{l}}
{\!\!\!\!-\!\;0.204\,{{\log }_{10}}({f_c})\! -\!6.28,1.5\,\rm{m}\!<\!{\emph{h}_\text{UT}}\!\le\!22.5\,\rm{m}}\\
{\!\!\!\!0.0965\,{{\log }_{{\rm{10}}}}({h_\text{UT}})\!-\!7.503,22.5\,\rm{m}\!<\!{\emph{h}_\text{UT}}\!\le\!300\,\rm{m}}\\
{{\!\!\!\!\rm{-7}}{\rm{.21}},\,\,{\rm{when\,the\,elevation\,angle\,of\,the\,link\,is\,}}{{10}^ \circ }}
\end{array}} \right.
\end{equation}
where $\text{lgDS}_{\mu ,{f_c}}={\log _{{\rm{10}}}}(\text{DS}_{{\mu ,{f_c}}}{\rm{/1s}})$.

In summary, we consider spatial consistency, the influences of carrier frequency, as well as the altitude of UT and elevation angle of link when generating LSPs. Generations of other seven LSPs are similar to the calculation of delay spread.
\subsection{STF Cluster Evolution}
The STF non-stationarity of the 6GPCM results from two mechanisms,~i.e., the STF varing parameters and the birth-death processes of clusters in three axes. Here, the birth-death process of clusters means some previous clusters disappear and some new clusters appear along these three axes~\cite{Wu2018_G5GCM,Liu2019_HST}. The number of clusters at time instant $t$, i.e., $N(t)$, contains the number of survived clusters $N_\text{surv}(t)$ and the newly generated clusters $N_\text{new}(t)$, can be calculated as
\begin{equation}
N(t)=N_\text{surv}(t)+N_\text{new}(t)
\end{equation}
where $N_\text{surv}(t)$ is determined by the survival probability of clusters $P_\text{surv}$. We assume the generation~(birth) and recombination~(death) rates of clusters are $\lambda_G$ and $\lambda_R$~\cite{Bian2021_B5G}, respectively. To describe the birth-death process more accurately, two types of sampling intervals can be used. The first type is the channel sampling intervals, such as $\Delta r$ in space domain, $\Delta t$ in time domain, and $\Delta f$ in frequency domain, within which channel parameters should be updated continuously. The other type can be described by $\Delta r_{\text{BD}}$, $\Delta t_{\text{BD}}$, and $\Delta f_{\text{BD}}$, which are integral multiples of corresponding $\Delta r$, $\Delta t$, and $\Delta f$. Birth-death processes of clusters in the corresponding domain occur at intervals $\Delta r_{\text{BD}}$, $\Delta t_{\text{BD}}$, and $\Delta f_{\text{BD}}$. Clusters that contribute to the received power must be seen by at least one Tx antenna element and one Rx antenna element at the same time in the same frequency bin. Besides, our model assumes that the propagation link of the Tx to a first-bounce cluster and the propagation link of the corresponding last-bounce cluster to the Rx are independent of each other. Therefore, ${P_\text{surv}}\left( {\Delta r_{\text{BD}},\Delta t_{\text{BD}},\Delta f_{\text{BD}}} \right) $ can be written as
\begin{align}
\label{equ_BD_process}
\nonumber & {P_\text{surv}}\left( {\Delta r_{\text{BD}},\Delta t_{\text{BD}},\Delta f_{\text{BD}}} \right) \\
& = P_\text{surv}^T\left( {\Delta t_{\text{BD}},{\delta _p}} \right)P_\text{surv}^R\left( {\Delta t_{\text{BD}},{\delta _q}} \right){P_\text{surv}}\left( {\Delta f_{\text{BD}}} \right)
\end{align}
where
\begin{equation}
P_\text{surv}^T\left( {\Delta t_{\text{BD}},{\delta _p}} \right) = {e^{ - {\lambda _R}{\sqrt{\left( {{{\left( {\varepsilon _1^T} \right)}^2} + {{\left( {\varepsilon _2^T} \right)}^2} + 2\varepsilon _1^T\varepsilon _2^T\cos \left( {\alpha _A^T - \beta _A^T} \right)} \right)}}}}
\end{equation}
\begin{equation}
P_\text{surv}^R\left( {\Delta t_{\text{BD}},{\delta _q}} \right) = {e^{ - {\lambda _R}{\sqrt{\left( {{{\left( {\varepsilon _1^R} \right)}^2} + {{\left( {\varepsilon _2^R} \right)}^2} + 2\varepsilon _1^R\varepsilon _2^R\cos \left( {\alpha _A^R - \beta _A^R} \right)} \right)}}}}
\end{equation}
\begin{equation}
{P_\text{surv}}\left( {\Delta f_{\text{BD}}} \right) = {e^{ - {\lambda _R}\frac{{F\left( {\Delta f_{\text{BD}}} \right)}}{{D_c^F}}}}.
\end{equation}
Here, $\Delta r_{\text{BD}}=\left\{\delta _p, \delta _q\right\}$, where $\delta_p=(p-1)\delta_T$ and $\delta_q=(q-1)\delta_R$. Also, $\delta_p$ denotes the spatial separation between $A_p^T$ to $A_1^T$ and $\delta_q$ represents the spatial separation between $A_q^R$ to $A_1^R$. Moreover, $\varepsilon_1^T=\frac{{{\delta _p}\cos \beta _E^T}}{{D_c^A}}$ ($\varepsilon_1^R=\frac{{{\delta _q}\cos \beta _E^R}}{{D_c^A}}$) and $\varepsilon _2^T=\frac{{{v^T}{\rm{\Delta }}t_{\text{BD}}}}{{D_c^S}}$ ($\varepsilon _2^R=\frac{{{v^R}{\rm{\Delta }}t_{\text{BD}}}}{{D_c^S}}$) represent the position differences of Tx (Rx) antenna element on array axis and time axis, respectively. In addition, values of $D_c^S$, $D_c^A$, $F\left( {\Delta f_{\text{BD}}} \right)$, and $D_c^F$ can be determined by channel measurements~\cite{Wu2018_G5GCM,psur_F}, and~\cite{3GPP38.901}.

The number of newly generated clusters $N_\text{new}(t)$ follows the Poisson distribution with mean value $\text{E}\left( {{N_\text{new}}} \right)$, which can be expressed as
\begin{equation}
\text{E}\left( {{N_\text{new}}} \right) = \frac{{{\lambda _G}}}{{{\lambda _R}}}\left( {1 - {P_\text{surv}}\left( {\Delta r_{\text{BD}},\Delta t_{\text{BD}},\Delta f_{\text{BD}}} \right)} \right).
\end{equation}

Note that in vacuum tube UHST scenarios, waveguide effect~\cite{waveguide} should be considered, and the mean value of newly generated clusters $\text{E}_\text{UHST}\left( {{N_\text{new}}} \right)$ can be calculated as
\begin{align}
\label{equ_bd_total}
\text{E}_\text{UHST}\left( {{N_\text{new}}} \right) = \text{E}\left( {{N_\text{new}}} \right) \left( {1 - \frac{{{D_{qp}}\left( t \right)}}{D}} \right) \cdot \frac{{{\rho _s}}}{{{\rho _{{s_0}}}}}
\end{align}
where $\rho _{s_0}$ is the scattering coefficient when the roughness ${\sigma _h} = 0$. Calculations of scattering coefficient $\rho _{s}$ can be referred to ~\cite{Xu2021_UHST}.

\subsection{Generation of New Clusters}
\label{subsec:new_clusters}
For a new cluster generated at time instant $t_0$, parameters like position, delay, angle, and power need to be assigned.
\subsubsection{Position of the Ray}
We use ellipsoid Gaussian scattering distribution \cite{Bian2021_B5G}, i.e., distributions of scatterers in the $n$th cluster with center ($\bar{d}_n^X, \bar{\phi}_{A,n}^X, \bar{\phi}_{E,n}^X$) on three axes follow Gaussian distribution with standard deviation of $\sigma_{x}^X$, $\sigma_{y}^X$, and $\sigma_{z}^X$, respectively. After obtaining the positions of the scatterers and converting them into spherical coordinates, $\vec{C}_{m_n}^A(t_0)$ and $\vec{C}_{m_n}^Z(t_0)$ relative to $\vec{A}^T_1(t_0)$ and $\vec{A}^R_1(t_0)$ are expressed by $\vec{C}_{m_n}^A(t_0) = \left( d_{{m_n}}^T(t_0), \phi_{A,{m_n}}^T(t_0), \phi_{E,{m_n}}^T(t_0) \right)$ and $\vec{C}_{m_n}^Z(t_0) = \left( d_{{m_n}}^R(t_0), \phi_{A,{m_n}}^R(t_0), \phi_{E,{m_n}}^R(t_0) \right)$, where $d_{{m_n}}^X(t_0)$, $\phi_{A,{m_n}}^X(t_0)$, and $\phi_{E,{m_n}}^X(t_0)$ are the distance, azimuth angle, and elevation angle of the $m$th ray in the $n$th cluster at the Tx or Rx side. The superscript $X\in\{T,R\}$ denotes the Tx and the Rx, respectively.

\subsubsection{Delay of the Ray}
As is shown in Fig.~\ref{6gchannels}, in a multi-bounce path, $\tau_{qp,m_n}(t_0)=(d_{{m_n}}^T(t_0)+d_{{m_n}}^R(t_0))/c+\tilde{\tau}_{m_n}(t_0)$, where $\tilde{\tau}_{m_n}$ is the time delay of virtual link between $\vec{C}^A_{m_n}$ and $\vec{C}^Z_{m_n}$, and $\tilde{\tau}_{m_n}=\tilde{d}_{m_n}/c+\tau_\text{link}$, where $\tilde{d}_{m_n}$ is distance between $\vec{C}^A_{m_n}$ and $\vec{C}^Z_{m_n}$, and $\tau_\text{link}$ is a non-negative variable which is assumed to be exponentially distributed.

\subsubsection{STF-Varying Ray Power}
\label{powers_rays}
In massive MIMO scenarios, the power ${P_{qp,{m_n},{f_c}}}\left( t\right)$ changes over both time and array axes. So we model the power as a lognormal process varying with time and a lognormal process varying along the antenna array, and the non normalized power can be calculated as
\begin{align}
\label{equ_powers_rays}
\nonumber & P_{qp,{m_n},{f_c}}^{\rm{'}}\left( {t} \right) = \\
& \underbrace {\text{exp}\left( { - {\tau _{qp,{m_n}}}\left( t \right)\frac{{{r_\tau } - 1}}{{{r_\tau }DS}}} \right){{10}^{ - \frac{{{Z_n}}}{{10}}}}}_{{\rm{\text{time}\,\text{domain}}}} \cdot \underbrace {{\xi _n}\left( {p,q} \right)}_{{\rm{\text{space}\,\text{domain}}}}.
\end{align}
Here, $DS$ is the delay spread mentioned before, $r_\tau$ denotes the delay distribution proportionality factor~\cite{3GPP38.901}, $Z_n$ is the per cluster shadowing term in dB following the Gaussian random distribution with zero mean. In addition, ${\xi _n}\left( {p,q} \right)$ is a two dimensional (2D) spatial lognormal process, which simulate the smooth power variations over antenna arrays~\cite{Bian2021_B5G}.

When discussing large bandwidth scenarios, the uncorrelated scattering assumption in frequency domain is not fulfilled~\cite{ultrawideband, Freq_nonstationary}, so that the influence of frequency on power should also be considered. we can multiply the power by $\left( {\frac{f}{{{f_c}}}} \right)^{{{\rm{\gamma }}_{{m_n}}}}$ in frequency domain to mimic the frequency-dependent property in large bandwidth scenarios, where $\gamma_{m_n}$ is the frequency-dependent factor \cite{frequency_factor}.

Then the final ray power ${P_{qp,{m_n},{f_c}}}\left( t\right)$ can be obtained using the normalization operation so that the sum of all cluster powers is equal to one. When a cluster is newly generated, the initial power can be obtained by substituting $\tau_{qp,m_n}(t_0)$ for $\tau_{qp,m_n}(t)$ in (\ref{equ_powers_rays}).

Furthermore, in multi-frequency channels, when generating SSPs, delays and angles of rays are the same at different carrier frequencies and the multi-frequency correlation is mainly reflected to the power~\cite{mmMagic}. Consequently, we generate delays and angles of rays using an anchor frequency, and detailed calculation process of powers can be referred to~\cite{3GPP38.901,Zhang2020_WCSP_MF}.
\subsection{Evolution of Survived Clusters}
For survived clusters, SSPs need to be updated at different time instants. For the trajectory segment at time instant $t_1$, i.e., next moment after clusters generation, the coordinate of ${A}^{T}_{p}$~is
\begin{align}
\vec{A}^{T}_{p}(t_1)={\vec{A}^{T}_{p}(t_0)}+{v^T}(t_1-t_0)\cdot\left[\begin{array}{c}
\cos\alpha_A^T \cdot \cos\alpha_E^T \\
\sin\alpha_A^T \cdot \cos\alpha_E^T \\
\sin\alpha_E^T
\end{array}\right]^{{T}}
\end{align}
where $\vec{A}^{T}_{p}(t_0)$ can be calculated as
\begin{align}
\vec{A}^{T}_{p}(t_0) = \vec{A}^{T}_{1}(t_0) + (p-1) \cdot \delta_{T} \cdot \left[\begin{array}{c}
\cos\beta_{A}^T \cdot \cos\beta_{E}^T \\
\sin\beta_{A}^T \cdot \cos\beta_{E}^T \\
\sin\beta_{E}^T
\end{array}\right]^{{T}}.
\end{align}
Meanwhile, $\vec{C}^A_{m_n}(t_1)$ can be obtained as
\begin{align}
\vec{C}^{A}_{m_n}(t_1)={\vec{C}^{A}_{m_n}(t_0)}+{v^{A_n}}(t_1-t_{0})\cdot\left[\begin{array}{c}
\cos\alpha_A^{A_n} \cdot \cos\alpha_E^{A_n} \\
\sin\alpha_A^{A_n} \cdot \cos\alpha_E^{A_n} \\
\sin\alpha_E^{A_n}
\end{array}\right]^{{T}}.
\end{align}
Then, the distance from ${A}^{T}_{p}$ to ${C}^A_{m_n}$ at time instant $t_1$ is calculated by $d^T_{p,m_n}(t_1) = \left \|\vec{C}^{A}_{m_n}(t_1) - \vec{A}^{T}_{p}(t_1)\right \|$, and we can get $d^R_{q,m_n}(t_1)$ using the same method. Therefore, $\tau_{qp,m_n}(t_1)=(d_{p,m_n}^T(t_1)+d_{q,m_n}^R(t_1))/c+\tilde{\tau}_{m_n}$. Besides, powers can be obtained according to (\ref{equ_powers_rays}) in Section~\ref{subsec:new_clusters}. We can get every $\tau_{qp,m_n}(t)$ and $P_{qp,m_n,f_c}(t)$ at time instant $t=t_2, t_3, ...$ , using geographical positions of Tx, Rx, and scatterers at previous time of the corresponding time instant.

In this way, based on spherical wave propagation mechanism, i.e., the angles of rays drift across the antenna array, SSPs for different antenna pairs can be obtained using geometry relations between Tx, Rx, and scatterers, which is closer to reality and increase the spatial resolution of the 6GPCM.
\begin{table*}[t!]\footnotesize
\setlength\tabcolsep{2.4pt}
\caption{Comparisons of the proposed 6GPCM, B5GCM, and typical standard 5G channel models.}
\label{tab_Compare_5Gs}
\begin{spacing}{1.125}
\begin{tabular}{|c|c|c|l|l|c|c|c|c|c|}
\hline
 \multicolumn{2}{|c|}{\text{Scenarios}  }                                                              & \text{\begin{tabular}[c]{@{}c@{}}Channel\\ Characteristics\end{tabular}}            & \multicolumn{2}{c|}{\text{\begin{tabular}[c]{@{}c@{}}Parameters and modeling methods of\\ the 6GPCM\end{tabular}}}                                                                                                               & \text{6GPCM}       & \text{\begin{tabular}[c]{@{}c@{}}3GPP\\ TR 38.901\end{tabular}} & \text{\begin{tabular}[c]{@{}c@{}}IMT-\\ 2020\end{tabular}} & \text{\begin{tabular}[c]{@{}c@{}}Qua-\\ DRiGa\end{tabular}} & \text{B5GCM}       \\ \hline

 \multirow{12}{*}{\begin{tabular}[c]{@{}c@{}} All\\spectra \end{tabular}} & \multirow{7}{*}{\begin{tabular}[c]{@{}c@{}}MmWave\\ /THz\\ Channel\end{tabular}} &  \begin{tabular}[c]{@{}l@{}}High resolution\end{tabular}     &  \multicolumn{2}{l|}{  \begin{tabular}[c]{@{}l@{}}Modeling the delay of rays ($\tau_{qp,m_n}(t)$)\end{tabular}  }                                                 & Yes                  & Yes                                                              & Yes                                                          & Yes                                                           & Yes                  \\ \cline{3-10}
                                                                                   & & \multirow{2}{*}{ \begin{tabular}[c]{@{}c@{}}Frequency domain\\non-stationarity \end{tabular}}      &  \multicolumn{2}{l|}{   \begin{tabular}[c]{@{}l@{}}1) Birth-death process in frequency domain \end{tabular} }& Yes                  & \textbf{\emph{No}}                                                               & \textbf{\emph{No}}                                                           & \textbf{\emph{No}}                                                            & \textbf{\emph{No}}                  \\ \cline{4-10}
                                                                                   & & & \multicolumn{2}{l|}{   \begin{tabular}[c]{@{}l@{}} 2) $P_{qp,m_{n},f_{c}}(t)$ varies  with frequency \\ \end{tabular} }& Yes                  & \textbf{\emph{No}}                                                               & \textbf{\emph{No}}                                                           & Yes                                                            & Yes                  \\ \cline{3-10}
                                                                                   & &\begin{tabular}[c]{@{}l@{}} Atmosphere\\absorption\end{tabular}                                                                        & \multicolumn{2}{l|}{\begin{tabular}[c]{@{}l@{}}1) Oxygen absorption at mmWave band ($AL$)\\ 2) Molecular absorption at THz band ($AL$)\end{tabular}}                                                                                                & Yes                  & Yes                                                              & Yes                                                          & \textbf{\emph{No}}                                                            & Yes                  \\ \cline{3-10}
                                                                                   & & Blockage effect                                                                       & \multicolumn{2}{l|}{\begin{tabular}[c]{@{}l@{}}Consider blockage effect ($BL$) on the received power\end{tabular}}                                                                                                          & Yes                  & Yes                                                              & Yes                                                          & \textbf{\emph{No}}                                                            & Yes                  \\ \cline{2-10}
&  \multirow{4}{*}{\begin{tabular}[c]{@{}c@{}} VLC Channel\end{tabular}} &  \begin{tabular}[c]{@{}c@{}}Negligible Doppler \\effect \& incoherent\\ light\end{tabular}     &  \multicolumn{2}{l|}{  \begin{tabular}[l]{@{}l@{}}Only consider powers ($P_{\!\!q,p_Hp_V\!,\lambda_T}^\text{LoS}\!(t)\!$, \!$P_{\!\!q,p_Hp_V\!,\lambda_T,m_n}^\text{NLoS}\!(t)\!$)\\and propagation delays ($\tau_{q,p_Hp_V}^\text{LoS}(t)$, $\tau_{q,p_Hp_V,m_n}^\text{NLoS}(t)$)\end{tabular}  }                                                 & Yes                  & \textbf{\emph{No}}                                                              & \textbf{\emph{No}}                                                          & \textbf{\emph{No}}                                                           & \textbf{\emph{No}}                  \\ \cline{3-10}
                                                                                   &  &                   \begin{tabular}[c]{@{}l@{}}3D rotational Rx\end{tabular}      &  \multicolumn{2}{l|}{   \begin{tabular}[c]{@{}l@{}} The elevation and azimuth angles of of the normal vector\\of Rx ($\beta^{R}_A(t)$ and $\beta^{R}_E(t)$) are time-variant\end{tabular} }& Yes                  & \textbf{\emph{No}}                                                               & \textbf{\emph{No}}                                                           & \textbf{\emph{No}}                                                            & \textbf{\emph{No}}                  \\ \cline{3-10}

                                                                                   & &                   \begin{tabular}[c]{@{}l@{}}Frequency domain\\
non-stationarity\end{tabular}      &  \multicolumn{2}{l|}{   \begin{tabular}[c]{@{}l@{}} Model the effective reflectance parameters of clusters\\ related to the wavelength range\end{tabular} }& Yes                  & \textbf{\emph{No}}                                                               & \textbf{\emph{No}}                                                           & \textbf{\emph{No}}                                                            & \textbf{\emph{No}}                  \\ \hline

 \multirow{12}{*}{\begin{tabular}[c]{@{}c@{}} Global-\\coverage\\scenarios\end{tabular}}  & \begin{tabular}[c]{@{}c@{}}LEO satellite\\ Channel\end{tabular}                                                                       & Ionosphere effect                                                                     & \multicolumn{2}{l|}{Faraday rotation matrix ($\mathbf{F}_{\mathrm{r}}$)}                                                                                                                                                      & Yes                  & \textbf{\emph{No}}                                                               & \textbf{\emph{No}}                                                           & \textbf{\emph{No}}                                                            & \textbf{\emph{No}}                   \\ \cline{2-10}

& \multirow{2}{*}{\begin{tabular}[c]{@{}l@{}}\\UAV Channel\end{tabular}}                                                      & 3D movement                                                                           & \multicolumn{2}{l|}{\begin{tabular}[c]{@{}l@{}}${{{\vec{v}}}^{T}}$, ${{{\vec{v}}}^{R}}$, ${{{\vec{v}}}^{A_n}}$, and ${{{\vec{v}}}^{Z_n}}$ all have elevation angles\end{tabular}}                                            & Yes                  & Yes                                                               & Yes                                                           & Yes                                                            & Yes                   \\ \cline{3-10}
                                                                                  & & \begin{tabular}[c]{@{}c@{}}LSPs relate to the\\ height\end{tabular}            & \multicolumn{2}{l|}{LSPs relate to the height of UAV $h_\text{UAV}$}                                                                                                                                                            & Yes                  & \textbf{\emph{No}}                                                               & \textbf{\emph{No}}                                                           & \textbf{\emph{No}}                                                            & \textbf{\emph{No}}                   \\ \cline{2-10}
&  \multirow{2}{*}{\begin{tabular}[c]{@{}c@{}}\\Maritime\\ Channel\end{tabular}}                                                      & \begin{tabular}[c]{@{}l@{}}Location\\ dependence\end{tabular}                                                                           & \multicolumn{2}{l|}{\begin{tabular}[c]{@{}l@{}} $h_{q p, f_{c}}^\text{LoS}(t, \tau)$, $h_{q p, f_{c}}^{\text{NLoS}_1}(t, \tau)$, and $h_{q p, f_{c}}^{\text{NLoS}_2}(t, \tau)$ will appear \\or disappear according to Tx/Rx locations in the channel\end{tabular}}                                            & Yes                  & \textbf{\emph{No}}                                                               & \textbf{\emph{No}}                                                           & \textbf{\emph{No}}                                                            & \textbf{\emph{No}}                   \\ \cline{3-10}
                                                                                  &  & \begin{tabular}[c]{@{}c@{}}Fluctuation of\\ sea waves\end{tabular}            & \multicolumn{2}{l|}{Pierson-Moskowitz (P-M) spectrum}                                                                                                                                                            & Yes                  & \textbf{\emph{No}}                                                               & \textbf{\emph{No}}                                                           & \textbf{\emph{No}}                                                           & \textbf{\emph{No}}                   \\ \hline

 \multirow{12}{*}{\begin{tabular}[c]{@{}c@{}} Full-\\application\\scenarios\end{tabular}}&  \multirow{2}{*}{\begin{tabular}[c]{@{}c@{}}\\V2V Channel\end{tabular}}                                                      & \begin{tabular}[c]{@{}c@{}}Arbitrary\\ trajectory\end{tabular}                        & \multicolumn{2}{l|}{\multirow{2}{*}{\begin{tabular}[c]{@{}c@{}}\\${{{\vec{v}}}^{T}}$, ${{{\vec{v}}}^{R}}$, ${{{\vec{v}}}^{A_n}}$, and ${{{\vec{v}}}^{Z_n}}$ are time-variant\end{tabular}}}                                                                              & Yes                  & Yes                                                               & Yes                                                           & Yes                                                            & Yes                  \\ \cline{3-3} \cline{6-10}
                                                                                  &  & \begin{tabular}[c]{@{}c@{}}Multi-mobility\\ property\end{tabular}                                                                          & \multicolumn{2}{l|}{}                                                                                                                                                                                                          & Yes   & \textbf{\emph{No}}  & \textbf{\emph{No}} & \textbf{\emph{No}}                                                                                                                                                            & Yes                  \\ \cline{2-10}

&  \multirow{4}{*}{\begin{tabular}[c]{@{}c@{}}(U)HST\\ Channel\end{tabular}}         & {\begin{tabular}[c]{@{}c@{}}Large Doppler\\ shift/spread\end{tabular}} & \multicolumn{2}{l|}{{$\nu_{qp,m_{n},f_{c}}(t)$ is time-variant}}                                                                                                                                                 & {Yes} & {Yes}                                             & {Yes}                                         & {Yes}                                          & {Yes} \\ \cline{3-10}
                                                                                  &  & \begin{tabular}[c]{@{}c@{}}Time domain \\non-stationarity\end{tabular}               & \multicolumn{2}{l|}{\begin{tabular}[c]{@{}l@{}}1) Cluster birth-death in time domain\\ 2) Channel parameters are time-variant\end{tabular}}                                                                                  & Yes                  & \textbf{\emph{No}}                                                              & \textbf{\emph{No}}                                                          & Yes                                                           & Yes                  \\ \cline{3-10}
                                                                                  &  & Waveguide effect                                                                      & \multicolumn{2}{l|}{\begin{tabular}[c]{@{}l@{}}$N_{qp}(t)$ relates to waveguide effects in different positions\end{tabular}}                                                                                                & Yes                  & \textbf{\emph{No}}                                                               & \textbf{\emph{No}}                                                           & \textbf{\emph{No}}                                                            & \textbf{\emph{No}}                   \\ \cline{2-10}

&  \multirow{2}{*}{\begin{tabular}[c]{@{}c@{}}Massive\\ MIMO\\ Channel\end{tabular}} & \begin{tabular}[c]{@{}c@{}}Spherical\\ wavefront\end{tabular}                         & \multicolumn{2}{l|}{$d^T_{p,m_n}$/$d^R_{q,m_n}$ is related to the AoD/AoA}                                                                                                                                                    & Yes                  & \textbf{\emph{No}}                                                               & \textbf{\emph{No}}                                                           & Yes                                                           & Yes                  \\ \cline{3-10}
&  & \begin{tabular}[c]{@{}c@{}}Space domain \\non-stationarity \end{tabular}          & \multicolumn{2}{l|}{\begin{tabular}[c]{@{}l@{}}1) Cluster birth-death in array domain\\ 2) $P_{qp,m_{n},f_{c}}(t)$ varies along the array axis (${\xi _n}\left( {p,q} \right)$)\end{tabular}}                         & Yes                  & \textbf{\emph{No}}                                                              & \textbf{\emph{No}}                                                          & Yes                                                           & Yes                  \\ \cline{2-10}

&  \multirow{2}{*}{RIS Channel}                                                      & \begin{tabular}[c]{@{}c@{}}Cascaded\\ sub-channel\end{tabular}                        & \multicolumn{2}{l|}{\begin{tabular}[c]{@{}l@{}}Model three sub-channels $\textbf{H}_{\text{IR}}$, $\textbf{H}_{\text{TI}}$, and $\textbf{H}_{\text{TR}}$, respectively\end{tabular}}                                         & Yes                  & \textbf{\emph{No}}                                                               & \textbf{\emph{No}}                                                           & \textbf{\emph{No}}                                                           & \textbf{\emph{No}}                   \\ \cline{3-10}
                                                                                  &  & Phase shift matrix                                                                    & \multicolumn{2}{l|}{Introduce the phase shift matrix $\mathbf{\Phi}$}                                                                                                                                                         & Yes                  & \textbf{\emph{No}}                                                               & \textbf{\emph{No}}                                                           & \textbf{\emph{No}}                                                            & \textbf{\emph{No}}                   \\ \cline{2-10}
&  IIoT Channel                                                                      & \begin{tabular}[c]{@{}c@{}}Dense multipath\\components\end{tabular}                                                                       & \multicolumn{2}{l|}{\begin{tabular}[c]{@{}l@{}}Model for rays in clusters, like $P_{qp,m_n,f_c}(t)$ \\ and $\tau_{qp,m_n}(t)$\end{tabular}}                                                                                  & Yes                  & \textbf{\emph{No}}                                                              & \textbf{\emph{No}}                                                          & \textbf{\emph{No}}                                                            & \textbf{\emph{No}}                   \\ \hline

\multirow{3}{*}{Common}&  \multicolumn{2}{c|}{Spatial consistency}                                                                                                                                 & \multicolumn{2}{l|}{Spatial correlated parameters}                                                                                                                                                                            & Yes                  & Yes                                                              & Yes                                                          & Yes                                                           & Yes                   \\ \cline{2-10}

&  \multicolumn{2}{c|}{Multi-frequency correlation} & \multicolumn{2}{l|}{\begin{tabular}[c]{@{}l@{}}$PL$, $P_{qp,m_n,f_c}(t)$, delay spread, and angle spreads \\are frequency dependent\end{tabular}}                                                                         & Yes                  & Yes  & Yes                                                                                               & Yes                                                           & \textbf{\emph{No}}                   \\ \hline
\end{tabular}
\end{spacing}
\vspace{-0.4 cm}
\end{table*}

\subsection{Simplified Channel Models}
\label{subsection_simplifiedCM}
Parameters and modeling methods of the 6GPCM in depicting various channel characteristics can be found in Table~\ref{tab_Compare_5Gs}. Also, 6GPCM can easily be reduced to various simplified channel models by adjusting channel model parameters, as shown in Table~\ref{Simp_6GPCM}. Specially, in scenarios employing RIS, we divide the channel into three sub-channels,~i.e., the channel between Tx and RIS, RIS and Rx, and Tx and Rx. The whole channel matrix is denoted as $\textbf{H}_\text{total}$, as
\begin{equation}
\label{total channel}
\begin{split}
    \textbf{H}_\text{total}&=(\textbf{H}_{\text{IR}}\mathbf{\Phi}\textbf{H}_{\text{TI}}+\textbf{H}_{\text{TR}})\textbf{f}
\end{split}
\end{equation}
where $\mathbf{\Phi}$ is the reflecting coefficients matrix of RIS and $\textbf{f}$ is the steering vector of Tx~\cite{Sun2021_IRS}. Calculation processes of the channel matrices of three sub-channels $\textbf{H}_{\text{IR}}$, $\textbf{H}_{\text{TI}}$, and $\textbf{H}_{\text{TR}}$ are the same as~(\ref{CIR_singleLink}), detailed information can be referred to~\cite{Sun2021_IRS}.
\begin{table*}[t!]\footnotesize
   \caption{Simplified models of the 6GPCM.}
   \label{Simp_6GPCM}
   \begin{spacing}{1.1}
    \begin{tabular}{|p{1.45cm}|p{2.05cm}|p{13cm}|}
      \hline
      6GPCM                 & Simplified model                 & \makecell[c]{Parameter Adjustments}              \\ \hline
      Multi-link                                &  Single-link             & $N_{T} = N_{R} = 1 $                \\ \hline
      Multi-frequency                 & Single-frequency                & \begin{tabular}[l]{@{}l@{}}1) $h_{qp,f_c}(t,\tau) = h_{qp}(t,\tau)$ \\2) Generate $P_{qp,m_{n},f_c}(t)$ without considering multi-frequency correlation\end{tabular} \\ \hline
                              &  \begin{tabular}[l]{@{}l@{}}Sub-6 GHz \\ (small bandwidth)\end{tabular}                         & \begin{tabular}[l]{@{}l@{}}1) $AL=1, BL=1$ \\ 2) Delays within a cluster are unresolvable: $M_n(t)=1$, $\tau_{qp,m_{n}}(t)=\tau_{qp,{n}}(t)$, $P_{qp,m_{n},f_c}(t)=P_{qp,n}(t)$\\ 3) Frequency domain stationarity: $\Delta f_{\text{BD}}=0, \gamma_{m_n}=0$ \end{tabular}                   \\ \cline{2-3}
      All spectra             & \begin{tabular}[l]{@{}l@{}}MmWave/THz+\\ultra-massive\\MIMO~\cite{THz_WangJ}\end{tabular}   & \begin{tabular}[l]{@{}l@{}}1) Single-link, single-frequency
                              \\ 2) $WE=1, \mu=1, \psi_{m_n}=0, M_n(t)=M_n$ \\3) Generate LSPs without considering spatial consistency \end{tabular}\\ \cline{2-3}

                              &  Indoor+VLC~\cite{Zhu_VLC}              & \begin{tabular}[l]{@{}l@{}}1) Single-link, single-frequency \\ 2) Tx is a $M_{p_H} \times M_{p_V}$ uniform planar LED array with spacings $\delta_H$ and $\delta_V$, $v^T=0$ \\ 3) $WE=1, AL=1, \mathbf{H_s}=1, M_n(t)=M_n$
                              \\ 4) $\Delta f_{\text{BD}}=0, \Delta t_{\text{BD}}=0 $, and 2D cluster evolution in space domain                                \end{tabular}  \\ \hline

                              &  LEO~\cite{LEO_faradayrotation}                         & \begin{tabular}[l]{@{}l@{}}1) Single-link, single-frequency \\ 2) $AL=1, BL=1, \mu=1, M_n(t)=M_n$
                              \\ 3) $\Delta f_{\text{BD}}=0, \gamma_{m_n}=0, {\xi _n}\left( {p,q} \right)=1$\\4) LSPs relate to the elevation angle of link \end{tabular}   \\ \cline{2-3}

      Global-coverage scenarios       &  \begin{tabular}[l]{@{}l@{}} UAV-to-\\ground~\cite{Chang2020_IoT}\end{tabular}  & \begin{tabular}[l]{@{}l@{}}1) Single-link, single-frequency
                              \\ 2) $AL=1, BL=1, WE=1, \mu=1, \psi_{m_n}=0, M_n(t)=M_n$ \\ 3) $\Delta f_{\text{BD}}=0, \gamma_{m_n}=0, {\xi _n}\left( {p,q} \right)=1$
                               \\4) LSPs relate to the $h_\text{UAV}$\end{tabular} \\ \cline{2-3}

                              & Maritime ship-to-ship~\cite{HeY_Maritime}             & \begin{tabular}[l]{@{}l@{}}1) Single-link, single-frequency \\ 2) $AL=1, BL=1, WE=1, \mu=1, \psi_{m_n}=0, M_n(t)=M_n$
                              \\ 3) $h_{q p, f_{c}}(t, \tau)=\sqrt{\frac{K_{R}(t)}{K_{R}(t)+1}} h_{q p, f_{c}}^\text{LoS}(t, \tau)+\sqrt{\frac{S_1}{K_{R}(t)+1}}h_{q p, f_{c}}^{\text{NLoS}_1}(t,\tau)+\sqrt{\frac{S_2}{K_{R}(t)+1}} h_{q p, f_{c}}^{\text{NLoS}_2}(t, \tau)$
                              \\ 4) $S_1+S_2=1, \Delta f_{\text{BD}}=0, \gamma_{m_n}=0, {\xi _n}\left( {p,q} \right)=1$\end{tabular}\\ \hline

                              &  V2V~\cite{BianJ_V2V19}                         & \begin{tabular}[l]{@{}l@{}}1) Single-link, single-frequency \\ 2) $AL=1, BL=1, WE=1, \mu=1, \psi_{m_n}=0, M_n(t)=M_n$
                              \\ 3) $\Delta f_{\text{BD}}=0, \gamma_{m_n}=0, {\xi _n}\left( {p,q} \right)=1$\end{tabular}   \\ \cline{2-3}
     Full-application scenarios  & MmWave+ UHST~\cite{Xu2021_UHST}   & \begin{tabular}[l]{@{}l@{}}
                               1) Single-link, single-frequency, clusters are distributed on the inner wall of the vacuum  tube
                               \\ 2)AL=1, BL=1, $WE=1, \mu=1, \psi_{m_n}=0, M_n(t)=M_n$  \\ 3) $v^{A_n}=v^{Z_n}=v^T=0$ \\ 4) $\Delta f_{\text{BD}}=0, \gamma_{m_n}=0, {\xi _n}\left( {p,q} \right)=1$ \end{tabular}  \\ \cline{2-3}

                              &  Ultra-massive MIMO~\cite{ZhengY_TVT22}  & \begin{tabular}[l]{@{}l@{}}1) Single-frequency \\ 2) $WE=1, \mu=1, \psi_{m_n}=0, M_n(t)=M_n$\end{tabular}  \\ \cline{2-3}
                              &  RIS~\cite{Sun2021_IRS}              & \begin{tabular}[l]{@{}l@{}}1) Single-link, single-frequency \\ 2) $AL=1, BL=1, WE=1, \mu=1, \psi_{m_n}=0, M_n(t)=M_n$ \\ 3) $\Delta f_{\text{BD}}=0, \gamma_{m_n}=0, \Delta t_{\text{BD}}=0$ \end{tabular}  \\ \cline{2-3}
                              &  IIoT~\cite{LiY_PIMRC}              & \begin{tabular}[l]{@{}l@{}}1) Single-link, single-frequency \\ 2) $WE=1, \mu=1, \psi_{m_n}=0$ \\ 3) $h_{q p, f_{c}}(t, \tau)=\sqrt{\frac{K_{R}(t)}{K_{R}(t)+1}} h_{q p, f_{c}}^\text{LoS}(t, \tau)+\sqrt{\frac{1}{K_{R}(t)+1}}(h_{q p, f_{c}}^{\text{NLoS}_\text{SMC}}(t, \tau)+h_{q p, f_{c}}^{\text{NLoS}_\text{DMC}}(t, \tau))$ \\ 4) $\Delta f_{\text{BD}}=0, \gamma_{m_n}=0 $\end{tabular}  \\ \hline

      Pervasive   & B5GCM~\cite{Bian2021_B5G}   & \begin{tabular}[l]{@{}l@{}}1) Single-link, single-frequency  \\ 2) $WE=1, \psi_{m_n}=0, M_n(t)=M_n$, $\Delta f_{\text{BD}}=0$
                              \\ 3) Generate LSPs without considering spatial consistency \end{tabular}  \\ \hline
    \end{tabular}
  \end{spacing}
  \vspace{-0.5 cm}
\end{table*}
\section{Statistical Properties of the Proposed Model}
\label{Sec_4}
\subsection{STF Correlation Function (STFCF)}
\label{subsec:Correlation_Function}
STFCF between ${H}_{qp}\left( t,f \right)$ and $H_{\tilde{q}\tilde{p}}\left( t+\Delta t,f+\Delta f \right)$ is defined as
\begin{align}
\nonumber  & {{R}_{qp,\tilde{q}\tilde{p}}}\left( t,f;\Delta r,\Delta t,\Delta f \right) \\
& =\text{E}\left[ {{H}_{qp}}\left( t,f \right)H_{\tilde{q}\tilde{p}}^{\text{*}}\left( t+\Delta t,f+\Delta f \right) \right].
\label{STFCorre}
\end{align}
Here, ${{H}_{qp}}\left( t,f \right)$ and $H_{\tilde{q}\tilde{p}}\left( t+\Delta t,f+\Delta f \right)$ are channel transfer functions (CTFs). Also, $\text{E}\left[\cdot\right]$ denotes the statistical average, $\left(\cdot\right)^{\text{*}}$ denotes the complex conjugation operation, $\Delta r$, $\Delta t$, $\Delta f$ are space, time, and frequency intervals, respectively. Besides, $\Delta r=\left\{ \Delta {{r}^{T}},\Delta {{r}^{R}} \right\}$, $\Delta {{r}^{T}}={{\delta }_{p}}-{{\delta }_{{\tilde{p}}}}$, $\Delta {{r}^{R}}={{\delta }_{q}}-{{\delta }_{{\tilde{q}}}}$.  The STFCF can be written as

\small
\begin{align}
\nonumber  & {{R}_{qp,\tilde{q}\tilde{p}}}\left( t,f;\Delta r,\Delta t,\Delta f \right) \\ \nonumber
 & =\text{E}\left[ \left( \sqrt{\frac{{{K}_{R}}\left( t \right)}{{{K}_{R}}\left( t \right)+1}}H_{qp,{{f}_{c}}}^{L}\left( t,f \right)+\sqrt{\frac{1}{{{K}_{R}}\left( t \right)+1}}H_{qp,{{f}_{c}}}^{N}\left( t,f \right) \right) \right. \\ \nonumber
 & \left( \sqrt{\frac{{{K}_{R}}\left( t+\Delta t \right)}{{{K}_{R}}\left( t+\Delta t \right)+1}}H_{\tilde{q}\tilde{p},{{f}_{c}}}^{\text{*}L}\left( t+\Delta t,f+\Delta f \right) \right. \\ \nonumber & \left. \left. +\sqrt{\frac{1}{{{K}_{R}}\left( t+\Delta t \right)+1}}H_{\tilde{q}\tilde{p},{{f}_{c}}}^{\text{*}N}\left( t+\Delta t,f+\Delta f \right) \right) \right] \\ \nonumber & =\sqrt{\frac{{{K}_{R}}\left( t \right)}{{{K}_{R}}\left( t \right)+1}\cdot \frac{{{K}_{R}}\left( t+\Delta t \right)}{{{K}_{R}}\left( t+\Delta t \right)+1}}R_{qp,\tilde{q}\tilde{p}}^{L}\left( t,f;\Delta r,\Delta t,\Delta f \right) \\ \nonumber
 & +\sqrt{\frac{1}{{{K}_{R}}\left( t \right)+1}\cdot \frac{1}{{{K}_{R}}\left( t+\Delta t \right)+1}}R_{qp,\tilde{q}\tilde{p}}^{N}\left( t,f;\Delta r,\Delta t,\Delta f \right) \\
\end{align}
\normalsize
where $R_{qp,\tilde{q}\tilde{p}}^{L}\left( t,f;\Delta r,\Delta t,\Delta f \right)$ is the STFCF of LoS component and $R_{qp,\tilde{q}\tilde{p}}^{N}\left( t,f;\Delta r,\Delta t,\Delta f \right)$ is the STFCF of NLoS component, respectively.
In addition, we can reduce STFCF to spatial cross-correlation function~(CCF) ${{R}_{qp,\tilde{q}\tilde{p}}}\left( t,f;\Delta r \right)$ by setting $\Delta t\text{ = 0}$ and $\Delta f\text{ = 0}$, to temporal auto-correlation function~(ACF) ${{R}_{qp}}\left( t,f;\Delta t \right)$ by setting $\Delta f\text{ = 0}$ and $\Delta r\text{ = 0}$,~i.e., $ p\text{ = }\tilde{p}$ and $q\text{ = }\tilde{q}$, and to frequency correlation function~(FCF) ${{R}_{qp}}\left( t,f;\Delta f \right)$ by setting $\Delta t\text{ = 0}$ and $\Delta r\text{ = 0}$.

\subsection{Delay Power Spectrum Density (PSD)}
Delay PSD, which is also named as power delay profile or multipath intensity profile, is the inverse Fourier transform of FCF ${{R}_{qp}}\left( t,f;\Delta f \right)$ with respect to (w.r.t.) $\Delta f$, and it can be expressed as
\begin{align}
{{S}_{qp}}\left( t,f;\tau\right)=\int {{R}_{qp}}\left( t,f;\Delta f\right){{e}^{j2\pi\tau\Delta f}}d\Delta f.
\label{PDP}
\end{align}
After derivation, the delay PSD is further written as
\begin{align}
{{S}_{qp}}\left( t,f;\tau\right)=\sum_{n=1}^{N_{qp}(t)} \sum_{m=1}^{M_{n}(t)} P_{qp, m_{n},f_c}(t,f) \delta\left(\tau-\tau_{qp,m_n}(t)\right).
\end{align}
Note that the delay PSD reflects the time-frequency dependent characteristic of the delay and power for the rays between the transmitting antenna element $A_p^T$ and the receiving antenna element $A_q^R$. The $P_{qp, m_{n},f_c}(t,f)$ is the power of corresponding ray and is affected by the time-frequency evolution of the clusters in time and frequency domains, and will further affect the delay PSD.
\subsection{Doppler PSD}
Doppler PSD is the Fourier transform of temporal ACF ${{R}_{qp}}\left( t,f;\Delta t \right)$ w.r.t. the time interval $\Delta t$, and can be written~as
\begin{align}
{{S}_{qp}}\left( t,f;\upsilon  \right)=\int {{R}_{qp}}\left( t,f;\Delta t \right){{e}^{-j2\pi \upsilon \Delta t}}d\Delta t
\end{align}
where $\upsilon$ is the Doppler frequency.
\subsection{Stationary Interval}
The stationary interval, which can evaluate the time-variation characteristics of channel, is the maximum time duration within which the channel can be considered as wide-sense stationary (WSS) channel. We can calculate the stationary interval using the method of local region of stationarity (LRS) \cite{stationarity_interval}. Firstly, the correlation coefficient of two delay PSDs can be calculate as
\begin{align}
{{R}_{\Lambda }}\left( t,f;\Delta t \right)\!=\!\frac{\int {{\text{S}}_{qp}}\left( t,f;\tau  \right){{\text{S}}_{qp}}\left( t+\Delta t,f;\tau  \right)d\tau }{\!\text{max}\!\left\{\! \int {\!{\text{S}}_{qp}}{{\left( t,f;\tau  \right)}^{2}}d\tau ,\int {\!{\text{S}}_{qp}}{{\left( t+\Delta t,f;\tau  \right)}^{2}}d\tau  \right\}\!}.
\end{align}
Then, the stationary interval can be obtained as the largest interval in which $ R_\Lambda \left( t,f;\Delta t \right)$ exceeds a given threshold ${c}_\text{thresh}$, which is usually set at 0.8 \cite{stationarity_interval},~i.e.,
\begin{align}
I\left( t,f \right)=\text{max} \left\{ {{\left. \Delta t \right|}{{{R}_{\Lambda }}\left( t,f;\Delta t \right) \ge c_\text{thresh}}} \right\}.
\end{align}
\subsection{Singular Value Spread (SVS)}
The singular value decomposition of the channel matrix can be obtained as
\begin{equation}\label{capacity}
\textbf{H}=\textbf{U}
\textbf{$\Sigma$}
\textbf{V}^\text{H}
\end{equation} where $\textbf{H}$ is the channel matrix, $\textbf{U}$ and $\textbf{V}$ represent unitary matrices, and matrix $\textbf{$\Sigma$}$ represent diagonal matrix with $N_T$ rows and $M_T$ columns. Furthermore, the SVS is defined as the ratio of the largest singular value to the smallest singular value and can be calculated as
\begin{equation}\label{svd}
\begin{aligned}
\kappa_{\text{svs}}=\frac{\underset{k}{\max}\,    {\sigma_k}}{\underset{k}{\min}\,
{\sigma_k}}
\end{aligned}
\end{equation} where $\sigma_k$ ($k$ = 1, 2, $\cdots$, $N_T$) are the singular values of the matrix $\textbf{$\Sigma$}$.

\subsection{Coherence Distance$ / $Time$ / $Bandwidth}
The coherence distance of an antenna array is the minimum antenna element spacing during which the spatial CCF equals to a given threshold $c_\text{thresh\underline{ }D} \in \left[ 0,1 \right]$ and the coherence distance at the Tx side can be obtained as
\begin{align}
{{D}_\text{c}}=\text{min}\left\{ {\Delta r}>0:{{R}_{qp,q\tilde{p}}}\left( t,f;{\Delta r} \right)=c_\text{thresh\underline{ }D} \right\}.
\end{align}
Coherence time is the minimum time difference during which the temporal ACF equals to a given threshold $c_\text{thresh\underline{ }T} \in \left[ 0,1 \right]$. The coherence time can be calculated as
\begin{align}
{{T}_\text{c}}=\text{min}\left\{ \Delta t>0:{{R}_{qp}}\left( t,f;\Delta t \right)=c_\text{thresh\underline{ }T} \right\}.
\end{align}
Similarly, coherence bandwidth, is the minimum frequency difference during which the FCF equals to a given threshold $c_\text{thresh\underline{ }B} \in \left[ 0,1 \right]$. The coherence bandwidth  can be expressed~as
\begin{align}
\label{equ_cb}
{{B}_\text{c}}=\text{min}\left\{ \Delta f>0:{{R}_{qp}}\left( t,f;\Delta f \right)=c_\text{thresh\underline{ }B} \right\}.
\end{align}

\subsection{Root Mean Square (RMS) Doppler Spread and RMS Delay Spread}
The RMS Doppler spread and RMS delay spread are utilized to measure the dispersion of signal in Doppler frequency domain and time delay domain. The RMS Doppler spread $\sigma _{\nu,qp}$ can be calculated as

\small
\begin{align}
{\sigma _{\nu,qp}}\left( t \right)=\sqrt{\left( \text{E}\left[ {\nu_{qp,{{m}_{n}}}}{{\left( t \right)}^{2}} \right]-\text{E}{{\left[ {\nu_{qp,{{m}_{n}}}}\left( t \right) \right]}^{2}} \right)}
\label{RMS_DS}
\end{align}
\normalsize
where $\nu_{qp, m_{n}}(t)$ is the Doppler frequency shift, which is caused by the movements of Tx, Rx, and scatterers.
Meanwhile, the RMS delay spread $\sigma _{\tau,qp}\left( t \right)$ can be calculated by replacing $\nu_{qp,m_n}$ with $\tau_{qp,m_n}$ in~(\ref{RMS_DS}), as

\small
\begin{align}
\sigma _{\tau,qp}\left( t \right)=\sqrt{\left( \text{E}\left[ {{\tau }_{qp,{{m}_{n}}}}{{\left( t \right)}^{2}} \right]-\text{E}{{\left[ {{\tau }_{qp,{{m}_{n}}}}\left( t \right) \right]}^{2}} \right)}.
\end{align}
\normalsize
\section{Results and Analysis}
\label{Sec_5}
Statistical properties of the 6GPCM for different frequency bands and scenarios are simulated and analyzed in this section. In the simulation, model parameters are chosen according to the minimum mean square error~(MMSE) method when there are corresponding measurement data. In particular, LSFs are ignored in RF-based channels since our model mainly focuses on SSFs. Moreover, the antenna arrays at both Tx and Rx are assumed to be omnidirectional ULA with relative spacing $\delta_T =\delta_R = \lambda /2$, where $\lambda$ is the wavelength. Parameters needed in LSPs generation are referred to Section~\ref{subsec:LSPs}. Unless otherwise noted, the other related parameters are listed as: $f_c=2.6$~GHz, $\beta_A^T = \pi/6$, $\beta_E^T=0$, $D=100$~m, $\sigma_{x}=3$~m, $\sigma_{y}=5$~m, $\sigma_{z}=4$~m, $\lambda_R=4$/m, $\lambda_G=80$/m, $M_T=128$, $M_R=1$, $\mu=1$, $\psi_{m_n}=0$.

\begin{figure}[t]
\centering\includegraphics[width=0.98\columnwidth]{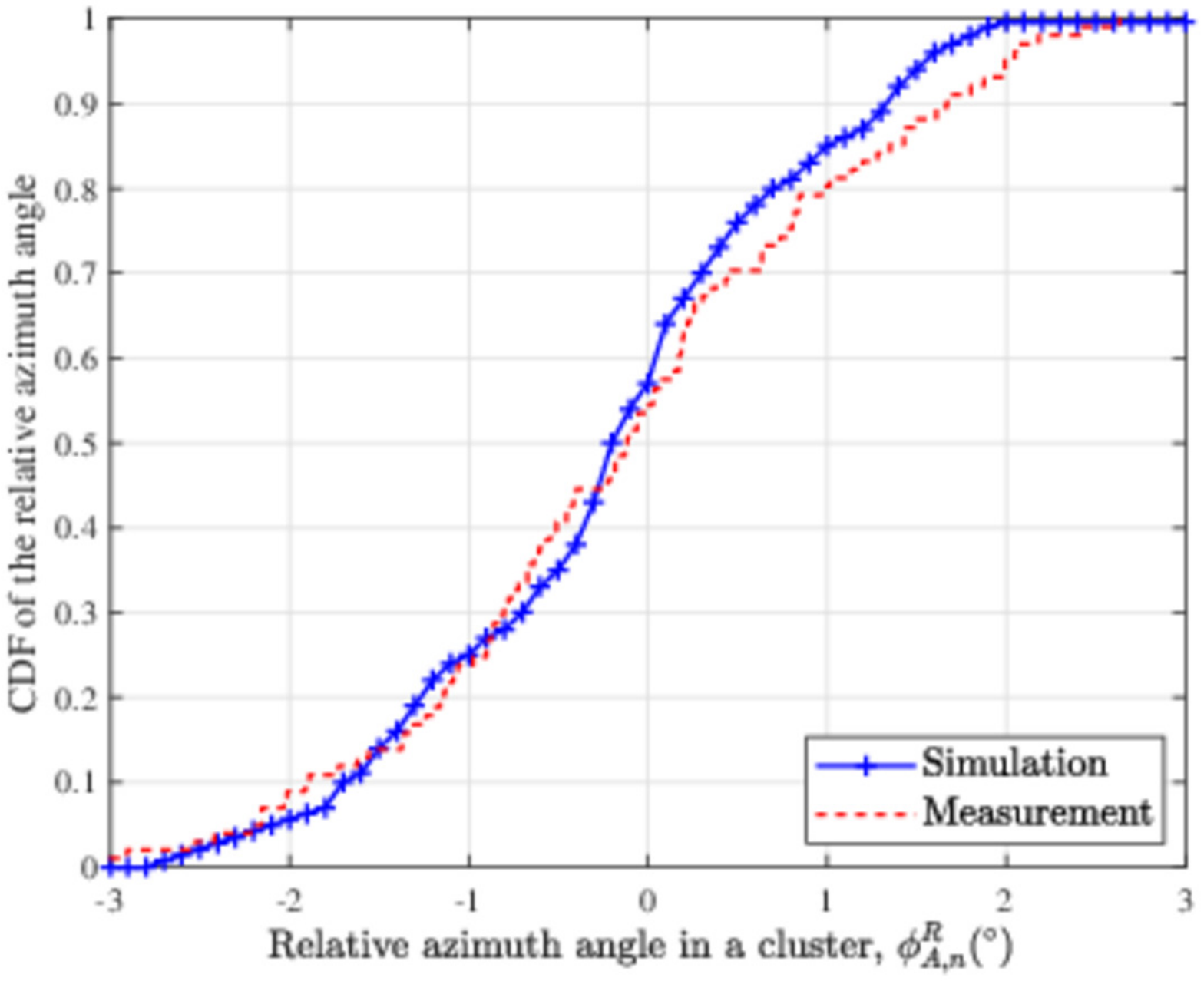}
\centering\caption{Cumulative distribution functions (CDFs) of the relative azimuth angles in a cluster in a small indoor scenario at THz band ($f_c$ = 300~GHz, $\sigma_{y}^R=1.4$~m, $v^{{T}}$ = $v^{{R}}$ = 0~m/s, $D$ = 2.8~m, $M_R = M_T = 1$, other parameters are shown in Table~\ref{tab_parameters}).}
\label{fig_THz}
\end{figure}

In terms of all spectra, we mainly analyzed characteristics of channels at THz and VLC bands as examples. Since the diffuse scattering in THz channel is mainly reflected in the cluster level angle spread~\cite{THz_WangJ}, we simulate the average relative AAoAs of rays in a cluster and compare them with the measurement data~\cite{THzMeas} in Fig.~\ref{fig_THz}. The measurements in~\cite{THzMeas} were conducted in a small indoor scenario with frequency from 275 GHz to 325 GHz and the distance between the Tx and the Rx is 2.8~m. We can observe that the result of the 6GPCM fits the measurement data well, which prove that our model can mimic the diffuse scattering characteristic of THz channels.

\begin{figure}[t]
\centering\includegraphics[width=0.98\columnwidth]{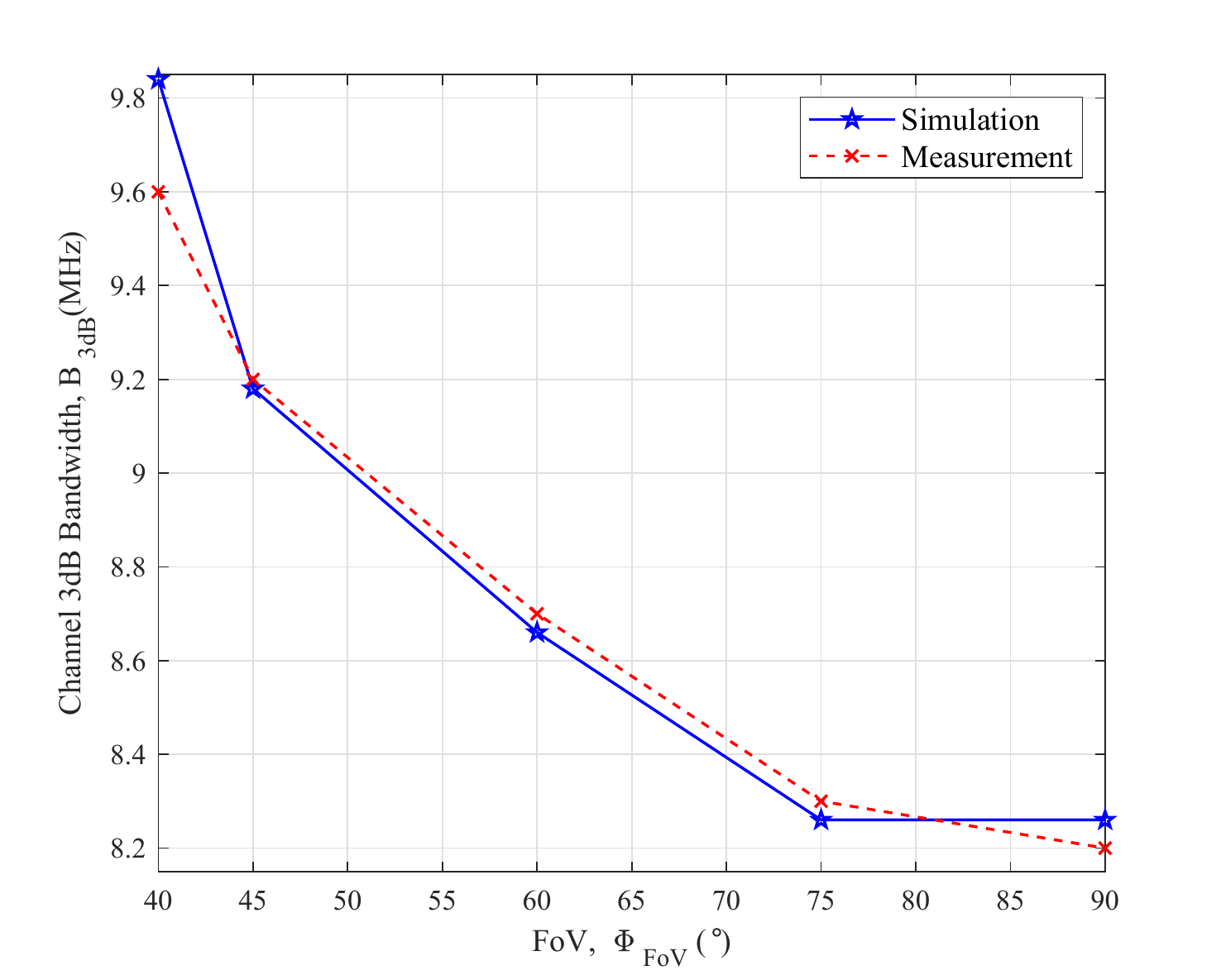}
\centering\caption{Channel 3-dB bandwidths with different FoVs of the proposed model and the measurement data in the VLC scenario ($\lambda=445$~nm, $D = 2.6345$ m, $\sigma_{x}^T=3.422$~m, $\sigma_{y}^T=2.691$~m, $\sigma_{z}^T=3.719$~m, $N(t_0)=10, M_n(t_0)=150$).}
\label{fig_VLC}
\end{figure}

For VLC communication systems, channel 3-dB bandwidths with different field of views (FoVs) of Rx are simulated and compared with the measurement data~\cite{vlcMeas} in Fig.~\ref{fig_VLC}. The channel measurement campaign was conducted in a typical indoor room using a blue-light VLC system with wavelength of 450 nm. The simulation result of our model and the measurement data can be found a good fit and our results are consistent with~\cite{Zhu_VLC}, showing that the 6GPCM can well support this communication scenario.

\begin{figure}[t]
\centering\includegraphics[width=0.95\columnwidth]{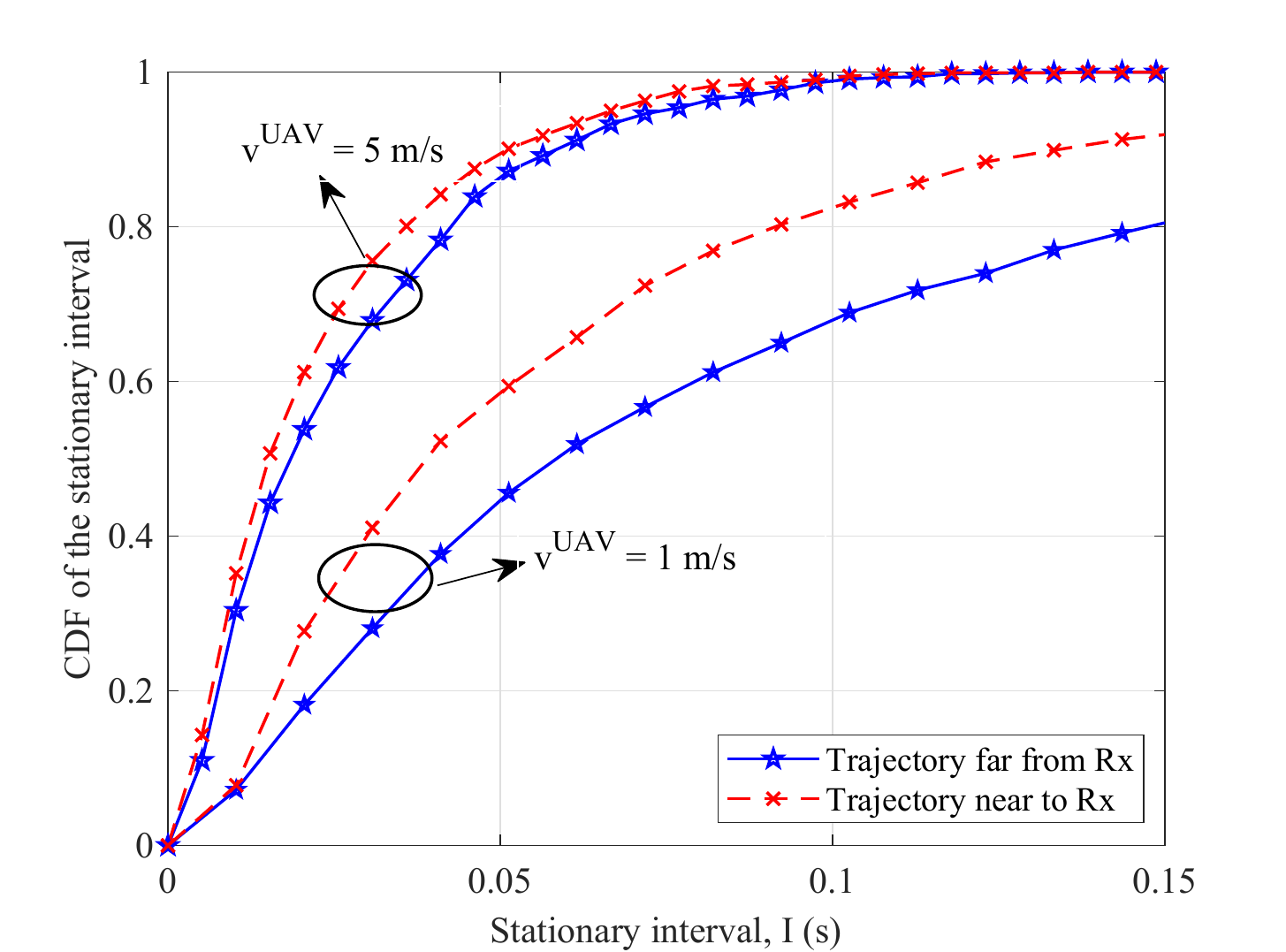}
\centering\caption{CDFs of stationary interval with different UAV trajectories ($f_c$~=~2.5~GHz, $v^{R}(t_0)$ = 0 m/s, $v^\text{UAV}(t_0)$ = 5 m/s, $a^{R} = a^\text{UAV} = 0 \,\text{m}/\text{s}^2$, $v^{A_n}=0 $~m/s, $v^{Z_n}=1.5$~m/s, other parameters are shown in Table~\ref{tab_parameters}).}
\label{fig_UAV_SI}
\end{figure}
In terms of global-coverage scenarios, we select UAV and maritime communication scenarios as examples. When the UAV flies with different trajectories and speeds, CDFs of the stationary interval are shown as Fig.~\ref{fig_UAV_SI}. In the simulation, the vertical distances from Rx to UAV are set to 50~m and 100~m. In UAV-to-ground channels, the faster the UAV moves, the smaller the value of stationary interval is. Also, the closer the UAV is to Rx, the smaller the value of stationary interval is. The reason for this possibility is that when UAV flies to the vicinity of Rx, the angular parameters will change dramatically, making the channel fluctuation more violent. The results are consistent with the corresponding results in~\cite{Chang2020_IoT}.
\begin{figure}[t]
\centering\includegraphics[width=1\columnwidth]{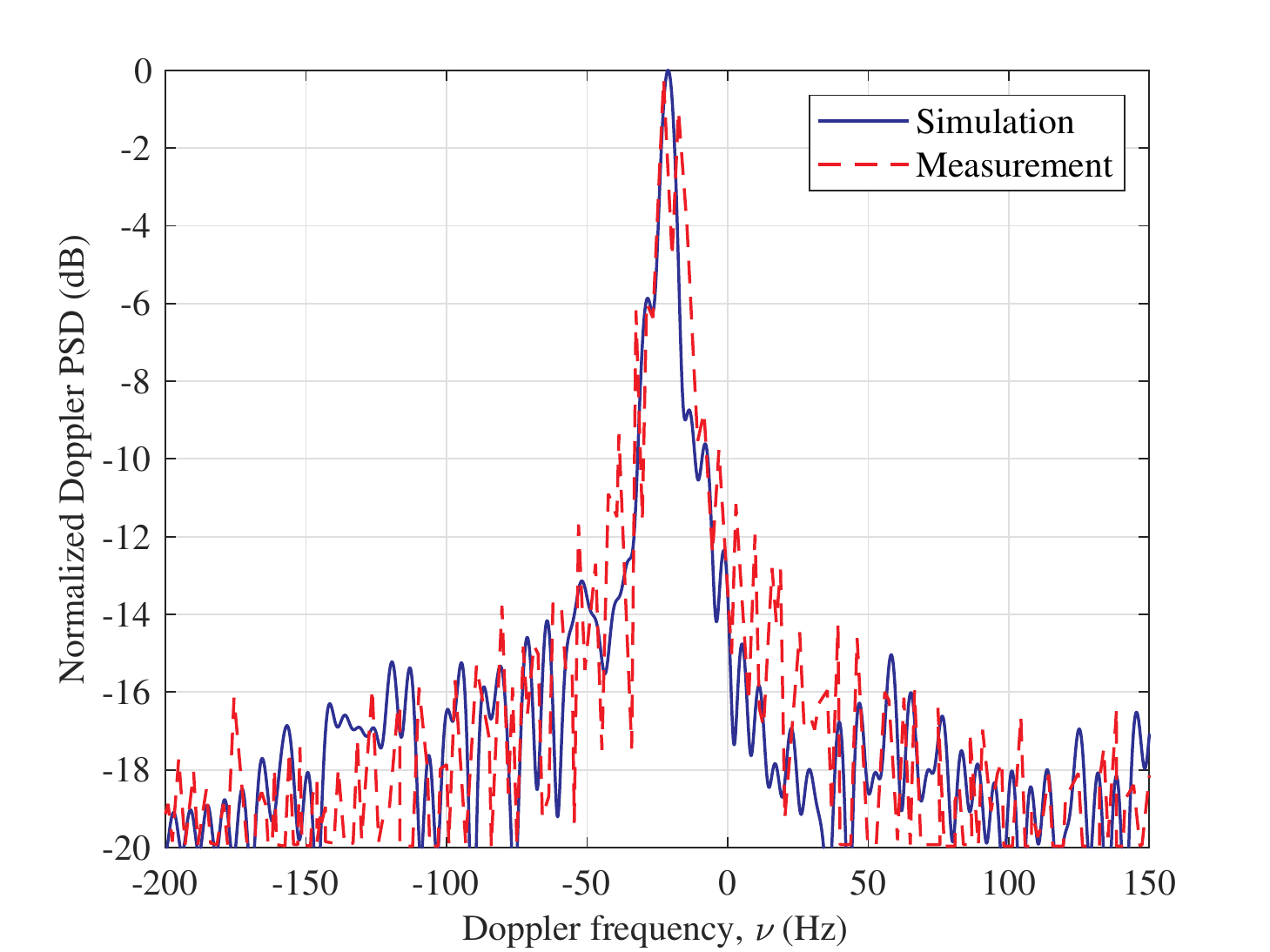}
\centering\caption{Normalized Doppler PSDs of 6GPCM and corresponding measurement data in the maritime ship-to-ship scenario ($f_c$~=~5.2~GHz, $v^{T}$ = $v^{R}$ = 7 m/s, $\beta^T_A=\pi/3, \beta^T_E=\pi/4,\beta^R_A=\pi/3, \beta^R_E=\pi/4$).}
\label{fig_Maritime}
\end{figure}

Fig.~\ref{fig_Maritime} gives the normalized Doppler PSD of 6GPCM and the corresponding measurement conducted in~\cite{Maritime_measure} when both ships were passing each other in a maritime ship-to-ship scenario. During the measurement, the speeds of the ships were chosen between 2 m/s and 7 m/s, 7 m/s is utilized in our simulation. Good agreement between simulation results and measurement data verifies the proposed channel model for its use in real maritime communication scenarios, which is consistent with~\cite{HeY_Maritime}.
\begin{figure}[t]
\centering\includegraphics[width=0.98\columnwidth]{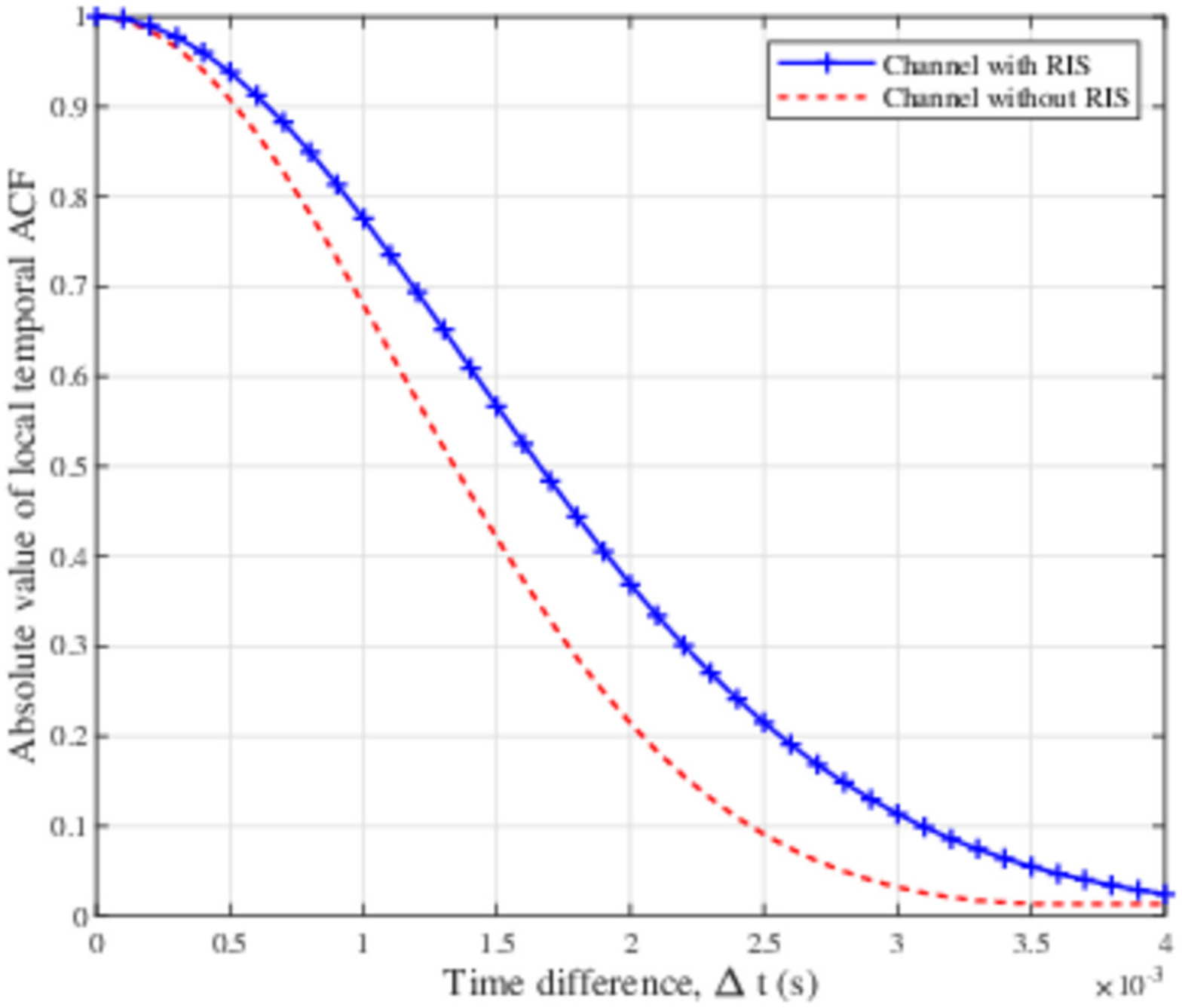}
\centering\caption{Temporal ACFs of the channels with RIS and without RIS ($f_c=62$~GHz, $v^{{T}}=v^{{R}}$ = 10~m/s, $D_{\text{TI}}=$ 100~m).}
\label{fig_IRS}
\end{figure}
\begin{figure}[t]
\centering\includegraphics[width=0.98\columnwidth]{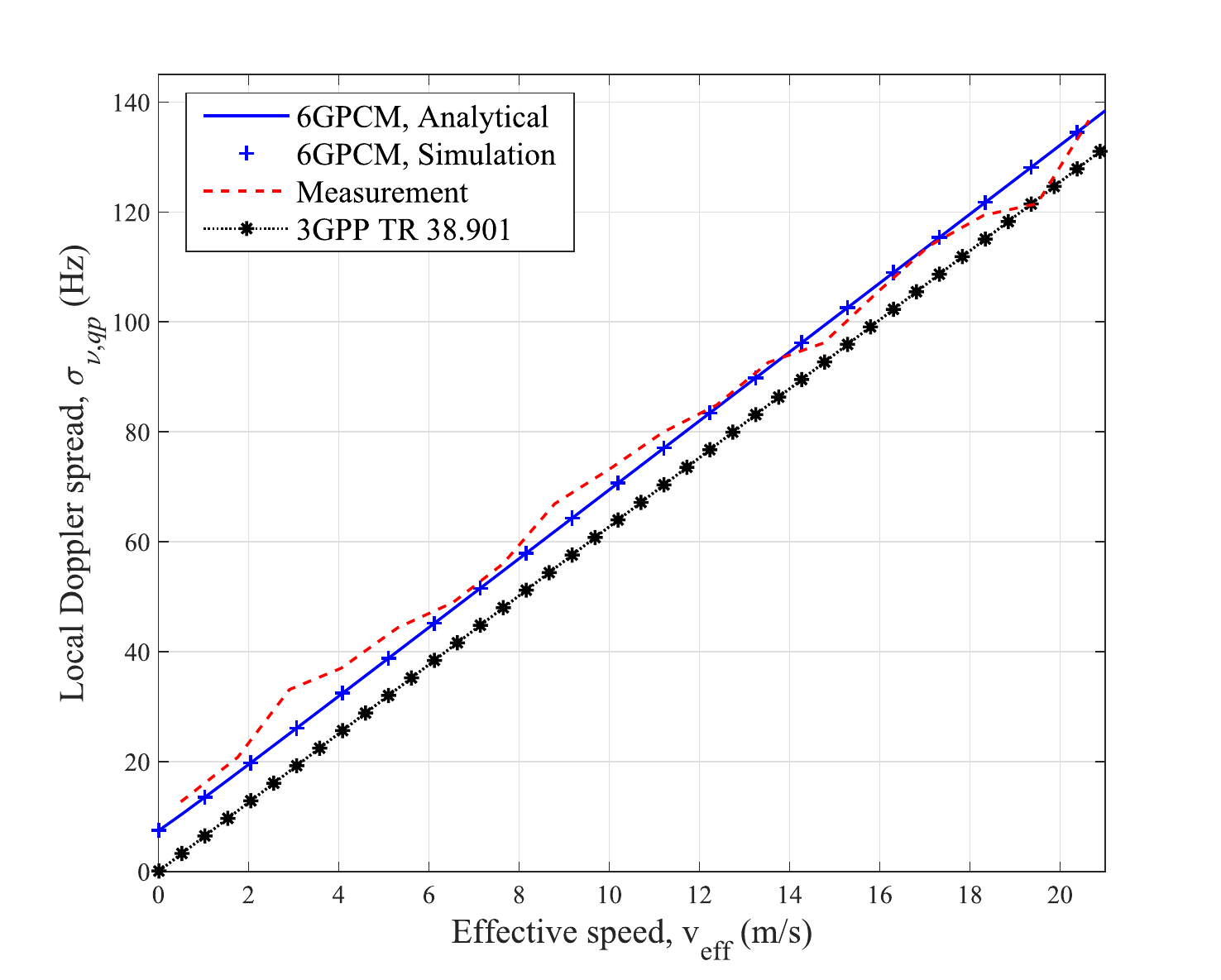}
\centering\caption{Doppler spreads of the 6GPCM and the measurement data~\cite{Cheng2008V2V_meas} in the V2V scenario ($f_c=5.9$~GHz, $\sigma_{x}=93$~m, $\sigma_{y}=103$~m, $\sigma_{z}=83$~m,  $\alpha^{T}=\alpha^{Z_n}=0$, $\alpha^{R}=\pi$, $v^{A_n}=0$~m/s, $v^{Z_n}=1.5$~m/s).}
\label{fig_Doppler_spread}
\end{figure}

In terms of full-application scenarios, we select RIS-based, V2V, vacuum tube UHST, ultra-massive MIMO, and IIoT scenarios as examples. In fact, the temporal ACF indicates the channel correlation with itself in the time domain, the larger the value is, the better the robustness of the channel is. Fig.~\ref{fig_IRS} shows the comparison of temporal ACFs of the channels with RIS and without RIS of same configurations. We can draw a conclusion that the application of RIS makes channel more robust, which is consistent with result in~\cite{Sun2021_IRS}.

The RMS Doppler spreads of the 6GPCM, 3GPP TR~38.901~\cite{3GPP38.901}, and the measurement data~\cite{Cheng2008V2V_meas} are compared in Fig.~\ref{fig_Doppler_spread}. The V2V channel measurements were conducted at 5.9~GHz in a suburban environment. The value of RMS Doppler spread shows a linear dependence relation with effective speed $v_\text{eff}=[{(v^T)}^2+{(v^R)}^2]^{\frac{1}{2}}$. The nonzero Doppler spread value when $v_\text{eff}=0$ comes from the motion of scatterers since both Tx and Rx are static. The theoretical result of the 6GPCM fits the simulation result well, illustrating the correctness of both mathematical derivations of the model and simulation results. Also, the analytical/simulation results fit measurement data well, demonstrating the validity of the proposed channel model in terms of RMS Doppler spread calculation. On the contrary, calculations from 3GPP TR 38.901~\cite{3GPP38.901} and other channel models assuming fixed scatterers do not agree with the corresponding measurement~data.

\begin{figure}[t]
\centering\includegraphics[width=0.98\columnwidth]{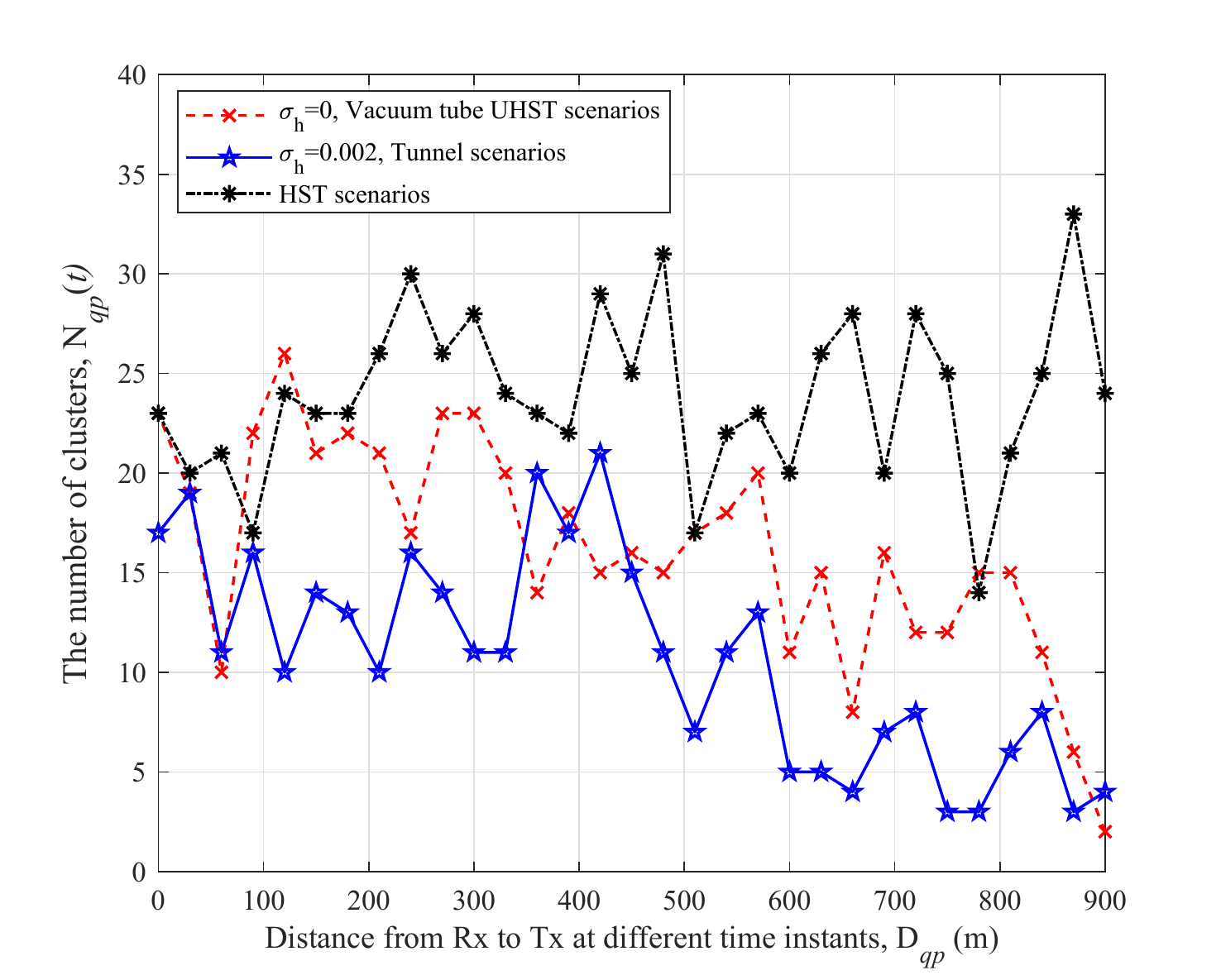}
\centering\caption{The number of clusters in three different (U)HST scenarios ($D=900$~m, $v^T=0$~km/h, $v^R=1080$~km/h, $f_c=58$~GHz, $\Delta t_{\text{BD}} =0.1$~s).}
\label{fig_UHST_NO_Clusters}
\end{figure}
Fig.~\ref{fig_UHST_NO_Clusters} illustrates the changing number of clusters w.r.t. the distance between Tx and Rx in three channels,~i.e., vacuum tube UHST channel, tunnel HST channel, and HST channel. The materials of tunnel wall and tube wall are generally reinforced concrete and low carbon steel, so we set ${\sigma _h} = 0.002$ and ${\sigma _h} = 0$ to simulate tunnel environment and vacuum tube environment, respectively~\cite{Wei2008_tunnel, Feng2019_vaccum}, and use \cite{Liu2019_HST} to model HST channel. As shown in this figure, the number of clusters in HST channel is extremely larger than the other two channels due to the narrow space of tunnel and vacuum tube, this phenomenon is in agreement with that in~\cite{Xu2021_UHST}.
\begin{figure}[t]
\setlength{\abovecaptionskip}{-0.1cm}
\centering\includegraphics[width=1.05\columnwidth,height=0.38\textwidth]{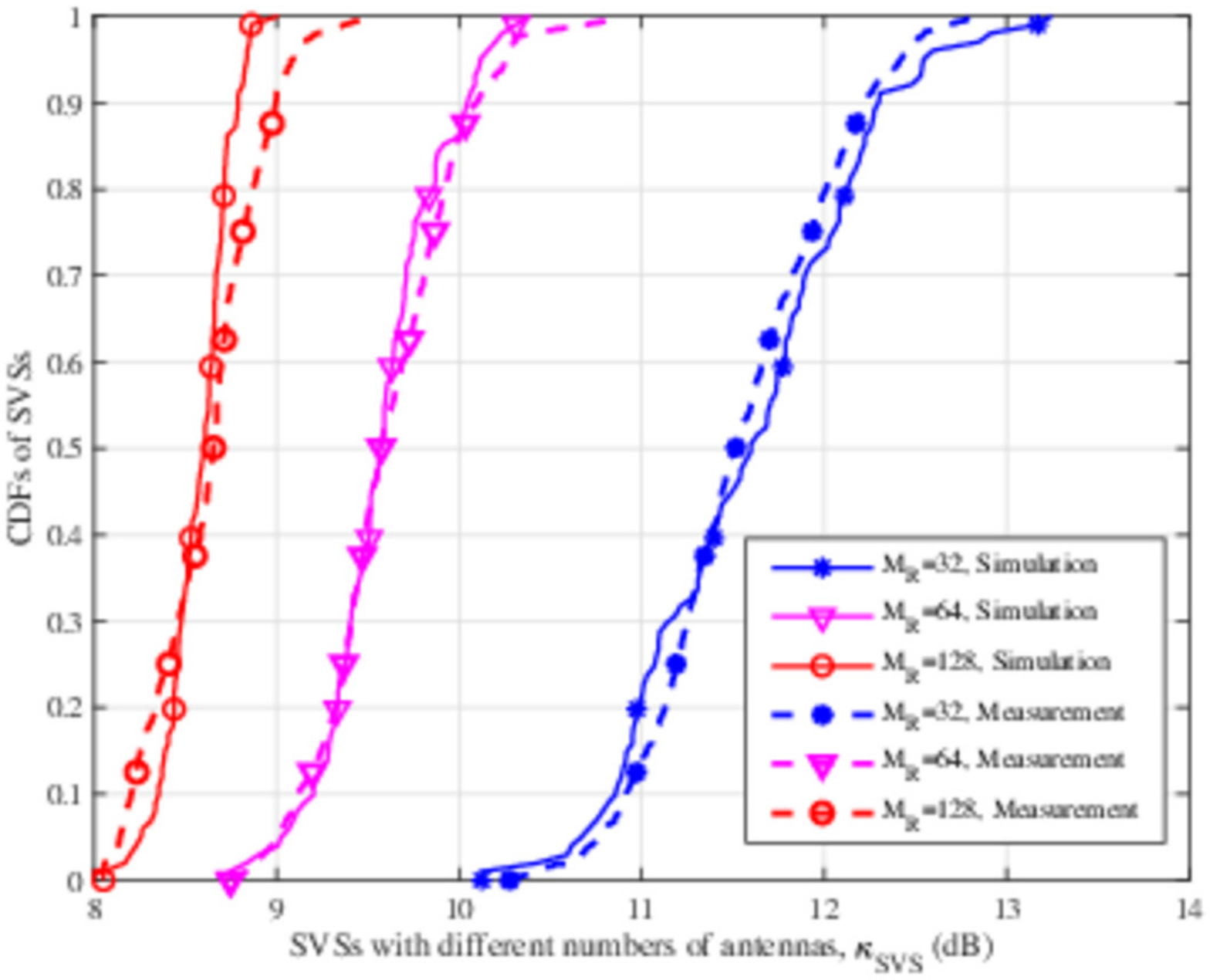}
\centering\caption{CDFs of SVSs with different $M_R$ in ultra-massive MIMO scenario (parameters in Table~\ref{tab_parameters}). }
\label{fig_SVS}
\end{figure}

\begin{table*}[t!]\footnotesize
   \caption{Configurations and parameter values of 6GPCM at different frequency bands and scenarios.}
   \label{tab_parameters}
   \begin{spacing}{1.3}
    \begin{tabular}{|l|l|l|l|l}
    \cline{1-4}
    \multicolumn{1}{|c|}{\multirow{2}{*}{Parameters}} & \multicolumn{3}{c|}{Parameter values at different frequency bands and scenarios}                                                               &  \\ \cline{2-4}
    \multicolumn{1}{|c|}{}   & indoor scenario at THz band~\cite{THz_WangJ}            & \multicolumn{1}{l|}{UAV-to-ground scenario~\cite{Chang2020_IoT}}  & \multicolumn{1}{l|}{ultra-massive MIMO scenario~\cite{ZhengY_TVT22}}   \\ \cline{1-4}
      $f_c$ (GHz)                                                               & 300 &  2.5     & 5.3        \\ \cline{1-4}
      Antenna type/$N_T$/$N_R$                                                  &  ULA/1/1         & ULA/1/1   & ULA/4/1          \\ \cline{1-4}
      $\vec{A}^T_1(t_0)$/$\vec{A}^R_1(t_0)$                                     & [0,0,0]/[2.8,0,0]  & [0,0,0]/[50,0,0]     & \begin{tabular}[l]{@{}l@{}}[$-50\sim$50, $-$37$\sim-$130,1.5] \\ /[0,0,20] \end{tabular}   \\ \cline{1-4}
      $M_T$/$\beta^T_A$/$\beta^T_E$/$\delta_T$                                  & 1/0/0/0.5$\lambda$   & 1/$\frac{\pi}{4}$/$\frac{\pi}{6}$/0.5$\lambda$   & 4/0/0/0.88$\lambda$  \\ \cline{1-4}
      $M_R$/$\beta^R_A$/$\beta^R_E$/$\delta_R$                                  &  1/0/0/0.5$\lambda$    & 1/$\frac{\pi}{4}$/$\frac{\pi}{4}$/0.5$\lambda$ & 128/0/$\frac{7\pi}{18}$/0.59$\lambda$ \\ \cline{1-4}
      Track of Tx/Rx                                                            & Static/Static         & Uniform linear motion/Static  & \begin{tabular}[l]{@{}l@{}}Uniform linear motion/Static\end{tabular}          \\ \cline{1-4}
      $v^T$(m/s)/$a^T$(m$^2$/s)/$\alpha^T_A$/$\alpha^T_E$                      & 0/0/0/0  & 5/0/$\frac{\pi}{2}$/0   & 0/0/$\frac{\pi}{2}$/0    \\ \cline{1-4}
      $v^R$(m/s)/$a^R$(m$^2$/s)/$\alpha^R_A$/$\alpha^R_E$                       & 0/0/0/0  & 0/0/0/0  & 0/0/0/0      \\ \cline{1-4}
      $\text{lgDS}={\log _{{\rm{10}}}}(\text{DS}{\rm{/1s}})$                     & $N(-7.72, 0.18), d_{\text{corr\_DS}}=6$ & \begin{tabular}[l]{@{}l@{}}$N(-0.31\text{log}_{10}(h_\text{UT})-6.845$, \\$0.7294\text{exp}(0.0014h_\text{UT})), d_{\text{corr\_DS}}=30$ \end{tabular}  & $N(-7.395, 0.1665), d_{\text{corr\_DS}}=30$       \\ \cline{1-4}
      $\text{lgASA}={\log _{{\rm{10}}}}(\text{ASA}{\rm{/1s}})$                  & $N(1.31, 0.855), d_{\text{corr\_ASA}}=8$  &    \begin{tabular}[l]{@{}l@{}}$N(-2.498\text{log}_{10}(h_\text{UT})-1.602$, \\$1.0389\text{exp}(0.0085h_\text{UT})), d_{\text{corr\_ASA}}=20$ \end{tabular} & $N(1.1392, 0.1069), d_{\text{corr\_ASA}}=20$   \\ \cline{1-4}
      $\text{lgASD}={\log _{{\rm{10}}}}(\text{ASD}{\rm{/1s}})$                  & $N(1.6, 0.18), d_{\text{corr\_ASD}}=7$  &  \begin{tabular}[l]{@{}l@{}}$N(-0.0135\text{log}_{10}(h_\text{UT})+1.345$, \\$1.0188\text{exp}(0.0001h_\text{UT})), d_{\text{corr\_ASD}}=50$ \end{tabular}  & $N(1.8699, 0.11), d_{\text{corr\_ASD}}=50$   \\ \cline{1-4}
      $\text{lgESA}={\log _{{\rm{10}}}}(\text{ESA}{\rm{/1s}})$                  &  $N(0.8, 0.165), d_{\text{corr\_ESA}}=4$   & \begin{tabular}[l]{@{}l@{}}$N(-0.289\text{log}_{10}(h_\text{UT})+0.225$, \\$0.9576\text{exp}(0.0018h_\text{UT})), d_{\text{corr\_ESA}}=50$ \end{tabular} & $N(1.2602, 0.16), d_{\text{corr\_ESA}}=50$    \\ \cline{1-4}
      $\text{lgESD}={\log _{{\rm{10}}}}(\text{ESD}{\rm{/1s}})$                  & $N(-1.31, 0.62), d_{\text{corr\_ESA}}=5$ & \begin{tabular}[l]{@{}l@{}}$N(-0.2975\text{log}_{10}(h_\text{UT})-0.5798$, \\ $1.0757\text{exp}(0.0059h_\text{UT})),d_{\text{corr\_ESD}}=50$\end{tabular}  & \begin{tabular}[l]{@{}l@{}}$N(\text{max}[-0.5, -2.1(\frac{d_\text{2D}}{1000})], 0.49)$,\\ $d_{\text{corr\_ESD}}=50$ \end{tabular}         \\ \cline{1-4}
      $SF$ (dB)                                                                  & $N(0, 3), d_{\text{corr\_SF}}=50$   & $N(0, 6), d_{\text{corr\_SF}}=50$  & $N(0, 6), d_{\text{corr\_SF}}=50$   \\ \cline{1-4}
      $\bar{d}_n^T$ (m)                                                        &$E(1), \text{min}_{d_n^T}=0.5$  & $E(20), \text{min}_{d_n^T}=10$  & $E(20), \text{min}_{d_n^T}=10$       \\ \cline{1-4}
      $\bar{d}_n^R$ (m)                                                       &$E(1), \text{min}_{d_n^T}=0.5$  & $E(15), \text{min}_{d_n^R}=10$    & $E(15), \text{min}_{d_n^R}=10$     \\ \cline{1-4}
      $\bar{\phi}_{E,n}^T$                                                    & $\text{ESD} \times N(0,1)\times\frac{\pi}{180}+\beta^T_E$  & $\text{ESD} \times N(0,1)\times\frac{\pi}{180}+\beta^T_E$ & $\text{ESD} \times N(0,1)\times\frac{\pi}{180}+\beta^T_E$      \\ \cline{1-4}
      $\bar{\phi}_{A,n}^T$                                                     & $\text{ASD} \times N(0,1)\times\frac{\pi}{180}+\beta^T_A$  & $\text{ASD} \times N(0,1)\times\frac{\pi}{180}+\beta^T_A$ & $\text{ASD} \times N(0,1)\times\frac{\pi}{180}+\beta^T_A$       \\ \cline{1-4}
      $\bar{\phi}_{E,n}^R$                                                         &$\text{ESA}\times N(0,1)\times\frac{\pi}{180}+\beta^R_E$ & $\text{ESA} \times N(0,1)\times\frac{\pi}{180}+\beta^R_E$ & $\text{ESA} \times N(0,1)\times\frac{\pi}{180}+\beta^R_E$       \\ \cline{1-4}
      $\bar{\phi}_{A,n}^R$                                                         & $\text{ASA} \times N(0,1)\times\frac{\pi}{180}+\beta^R_A$     & $\text{ASA} \times N(0,1)\times\frac{\pi}{180}+\beta^R_A$ & $\text{ASA} \times N(0,1)\times\frac{\pi}{180}+\beta^R_A$   \\ \cline{1-4}
      $\sigma^T_{x}$/$\sigma^T_{y}$/$\sigma^T_{z}$                           &$N(1.4, 0.1)$/$N(1.4, 0.1)$/$N(1, 0.1)$  & $N(5, 0.1)$/$N(5, 0.1)$/$N(5, 0.1)$  & $N(4, 0.1)$/$N(15, 0.1)$/$N(7, 0.1)$   \\ \cline{1-4}
      $\sigma^R_{x}$/$\sigma^R_{y}$/$\sigma^R_{z}$                           & $N(1.4, 0.1)$/$N(1.4, 0.1)$/$N(1, 0.1)$  & $N(5, 0.1)$/$N(5, 0.1)$/$N(5, 0.1)$   & $N(3.9, 0.1)$/$N(2, 0.1)$/$N(7, 0.1)$    \\ \cline{1-4}
      $\lambda_G$/$\lambda_R$ (m$^{-1}$)                                      & 0.8/0.04   &  20/1    & 20/1    \\ \cline{1-4}
      $D_c^S$ (m)/$D_c^A$ (m)/$D_c^F$ (GHz)                                     &  0/0/1      & 30/0/0  & 40/2/0     \\ \cline{1-4}
      $\tau_\text{link}$ (s)                                                   & \begin{tabular}[l]{@{}l@{}}$E(44.1\times10^{-9})$, \\ $\text{min}_{\tau_\text{link}}=27.3\times10^{-9}$\end{tabular}  & \begin{tabular}[l]{@{}l@{}}$E(44.1\times10^{-9})$, \\ $\text{min}_{\tau_\text{link}}=27.3\times10^{-9}$\end{tabular}  &\begin{tabular}[l]{@{}l@{}} $E(74.8\times10^{-9})$, \\ $\text{min}_{\tau_\text{link}}=14\times10^{-9}$\end{tabular}  \\ \cline{1-4}
      $r_\tau$/$\gamma_{m_n}$                                                   & 3/1.2 & 3/0  & 7.6096/0     \\ \cline{1-4}
    \end{tabular}
  \end{spacing}
  \vspace{-0.4 cm}
\end{table*}
Fig.~\ref{fig_SVS} shows the CDFs of SVSs of simulation results and measurement data in NLoS condition in ultra-massive MIMO scenario. The measurement was conducted by our group at outdoor UMa scenario in NLoS condition and there are 4 users with the distance of 1 m \cite{ZhengY_TVT22}. The detailed configuration and simulation parameters are shown in Table~\ref{tab_parameters}. In the simulation, we fixed $M_T =$~4 and increased the number of antennas at Rx side $M_R$ to 128 gradually. From the figure, we can see that the simulation results correspond to the measurement data well. Also, with the increase of $M_R$, the value of SVS gradually decreases, that means the channel vectors among different users become more and more orthogonal and channel hardening phenomenon is more and more obvious.

\begin{figure}[t]
\centering
\subfigure[LoS condition]{\includegraphics[width=0.49\textwidth]{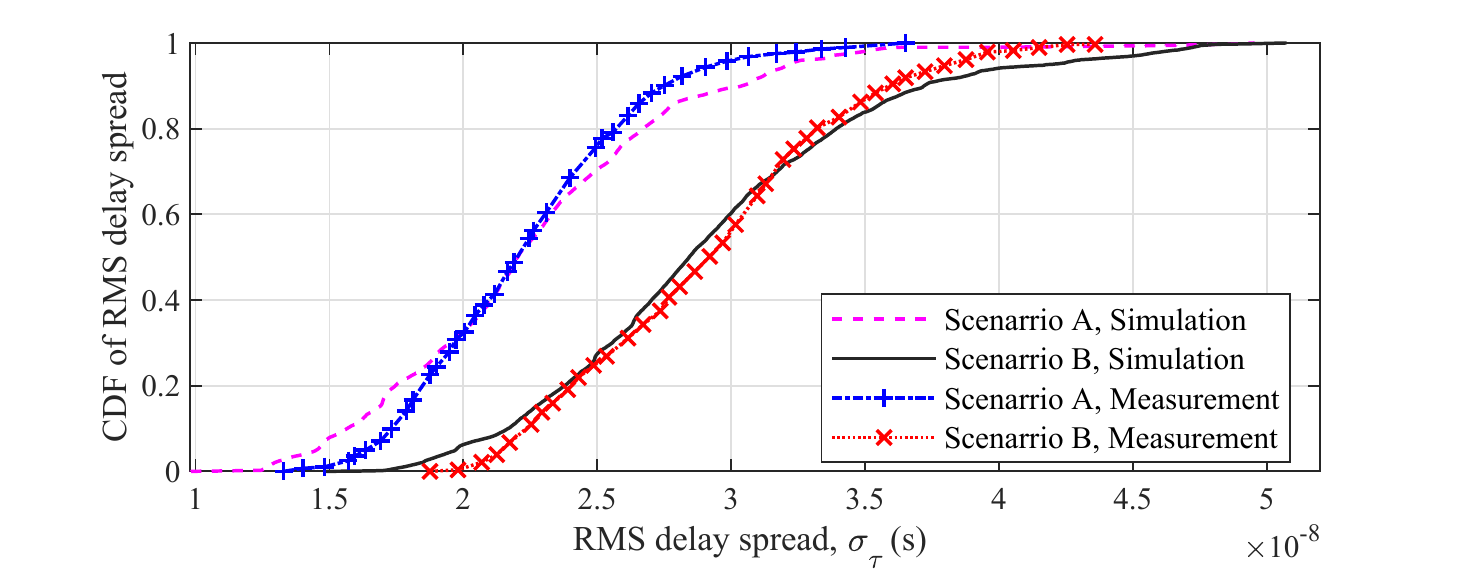}} \\
\vspace{-0.2 cm}
\subfigure[NLoS condition]{\includegraphics[width=0.49\textwidth]{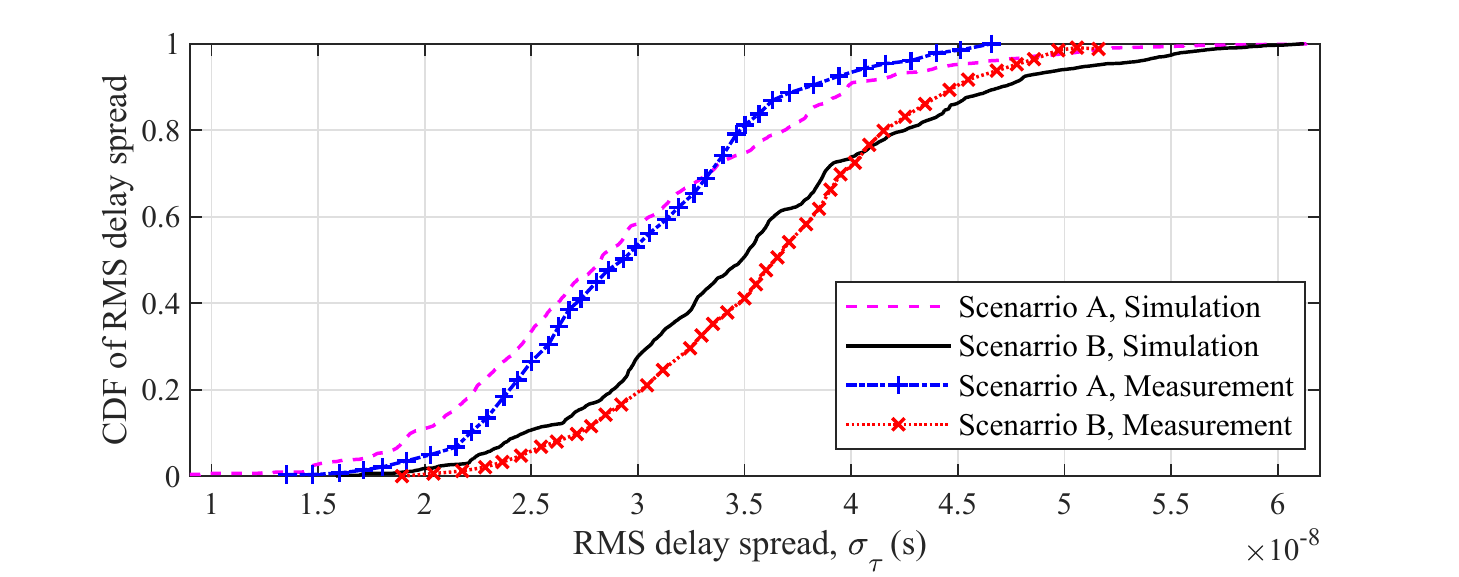}} \\
\caption{Simulation results and corresponding measurement data of CDFs of RMS delay spread in different IIoT scenarios ($N(t_0)=15$ in scenario A, $N(t_0)=20$ in scenario B, $f_c$ = 5.8 GHz, $M_R = M_T = 1$, $v^R$ = 0.4 m/s, $v^T = v^{A_n} = v^{Z_n} = $ 0 m/s).}
\label{fig_IIOT}
\end{figure}
There are four IIoT sub-scenarios on the basis of the antenna height and clutter density~\cite{3GPP38.901}. We choose measurement data of two sub-scenarios,~i.e., light clutter~(Scenario A) and heavy clutter~(Scenario B) environments, in LoS and NLoS conditions~\cite{Holfeld2016_IIOT}. Fig.~\ref{fig_IIOT} represents comparisons of simulations and measurements of CDFs of RMS delay spread at two scenarios in LoS and NLoS conditions. It can be seen that the simulation results can be well matched with the measurement results. Also, the delay spread in NLoS scenario is larger than that in LoS scenario, and delay spread in Scenario B is larger than that in Scenario A. The results are consistent with the corresponding results in~\cite{LiY_PIMRC}.
\section{Conclusions}
\label{sec_Conclusions}
In this paper, we have first proposed a pervasive wireless channel modeling theory and then applied it to a GBSM constructing a 6GPCM. The proposed 6GPCM has considered all important channel characteristics at all spectra, global-coverage scenarios, and full-application scenarios in the 6G wireless communication systems. Also, the 6GPCM can be simplified to standard 5G channel models by adjusting channel model parameters. In addition, key statistical properties of the 6GPCM have been derived, simulated, and compared with many channel measurements at specific frequency bands and scenarios, showing good fitting results. These have clearly proved the correctness of derivation and simulation results, accuracy, pervasiveness, and applicability of the proposed 6GPCM. It is expected that the proposed 6GPCM will play a fundamental supporting role for the future 6G channel model standardization and design, optimization, and performance evaluation of the 6G wireless communication systems.


\begin{IEEEbiography}[{\includegraphics[width=1.1in,height=1.3in,clip,keepaspectratio] {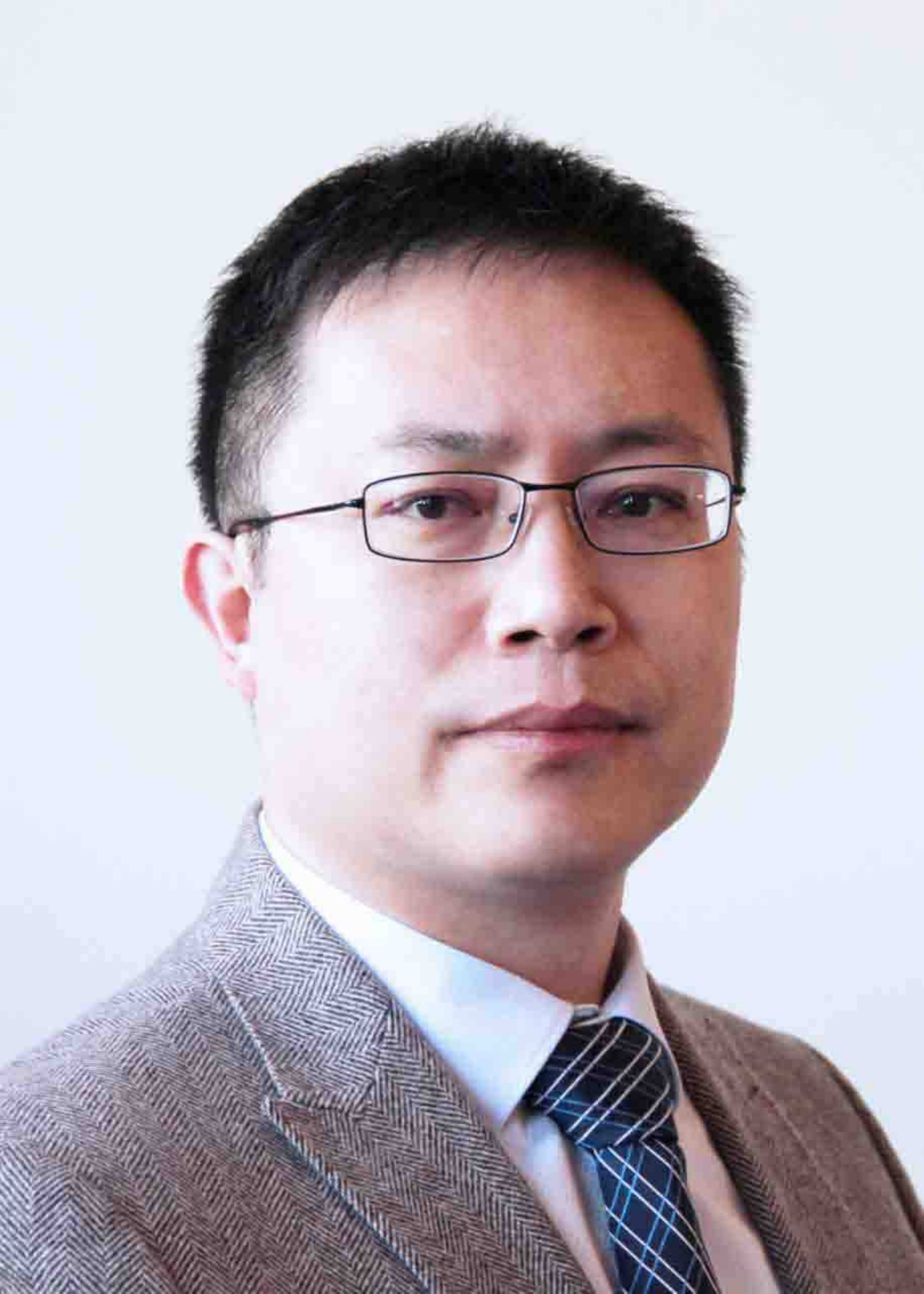}}]
{Cheng-Xiang Wang} (S'01-M'05-SM'08-F'17) received the B.Sc. and M.Eng. degrees in communication and information systems from Shandong University, China, in 1997 and 2000, respectively, and the Ph.D. degree in wireless communications from Aalborg University, Denmark, in 2004.

He was a Research Assistant with the Hamburg University of Technology, Hamburg, Germany, from 2000 to 2001, a Visiting Researcher with Siemens AG Mobile Phones, Munich, Germany, in 2004, and a Research Fellow with the University of Agder, Grimstad, Norway, from 2001 to 2005. He has been with Heriot-Watt University, Edinburgh, U.K., since 2005, where he was promoted to a Professor in 2011. In 2018, he joined Southeast University, Nanjing, China, as a Professor. He is also a part-time Professor with Purple Mountain Laboratories, Nanjing. He has authored 4 books, 3 book chapters, and more than 460 papers in refereed journals and conference proceedings, including 26 highly cited papers. He has also delivered 23 invited keynote speeches/talks and 10 tutorials in international conferences. His current research interests include wireless channel measurements and modeling, 6G wireless communication networks, and electromagnetic information theory.

Dr. Wang is a Member of the Academia Europaea (The Academy of Europe), a Fellow of the Royal Society of Edinburgh (FRSE), IEEE, IET, and China Institute of Communications (CIC), an IEEE Communications Society Distinguished Lecturer in 2019 and 2020, and a Highly-Cited Researcher recognized by Clarivate Analytics in 2017-2020. He is currently an Executive Editorial Committee Member of the IEEE TRANSACTIONS ON WIRELESS COMMUNICATIONS. He has served as an Editor for over ten international journals, including the IEEE TRANSACTIONS ON WIRELESS COMMUNICATIONS, from 2007 to 2009, the IEEE TRANSACTIONS ON VEHICULAR TECHNOLOGY, from 2011 to 2017, and the IEEE TRANSACTIONS ON COMMUNICATIONS, from 2015 to 2017. He was a Guest Editor of the IEEE JOURNAL ON SELECTED AREAS IN COMMUNICATIONS, Special Issue on Vehicular Communications and Networks (Lead Guest Editor), Special Issue on Spectrum and Energy Efficient Design of Wireless Communication Networks, and Special Issue on Airborne Communication Networks. He was also a Guest Editor for the IEEE TRANSACTIONS ON BIG DATA, Special Issue on Wireless Big Data, and is a Guest Editor for the IEEE TRANSACTIONS ON COGNITIVE COMMUNICATIONS AND NETWORKING, Special Issue on Intelligent Resource Management for 5G and Beyond. He has served as a TPC Member, a TPC Chair, and a General Chair for more than 80 international conferences. He received 14 Best Paper Awards from IEEE GLOBECOM 2010, IEEE ICCT 2011, ITST 2012, IEEE VTC 2013 Spring, IWCMC 2015, IWCMC 2016, IEEE/CIC ICCC 2016, WPMC 2016, WOCC 2019, IWCMC 2020, WCSP 2020, CSPS 2021, and WCSP 2021. Also, he received the 2020-2022 ``AI 2000 Most Influential Scholar Award Honourable Mention" in recognition of his outstanding and vibrant contributions in the field of Internet of Things.
\end{IEEEbiography}
\begin{IEEEbiography}[{\includegraphics[width=1.1in,height=1.3in,clip,keepaspectratio] {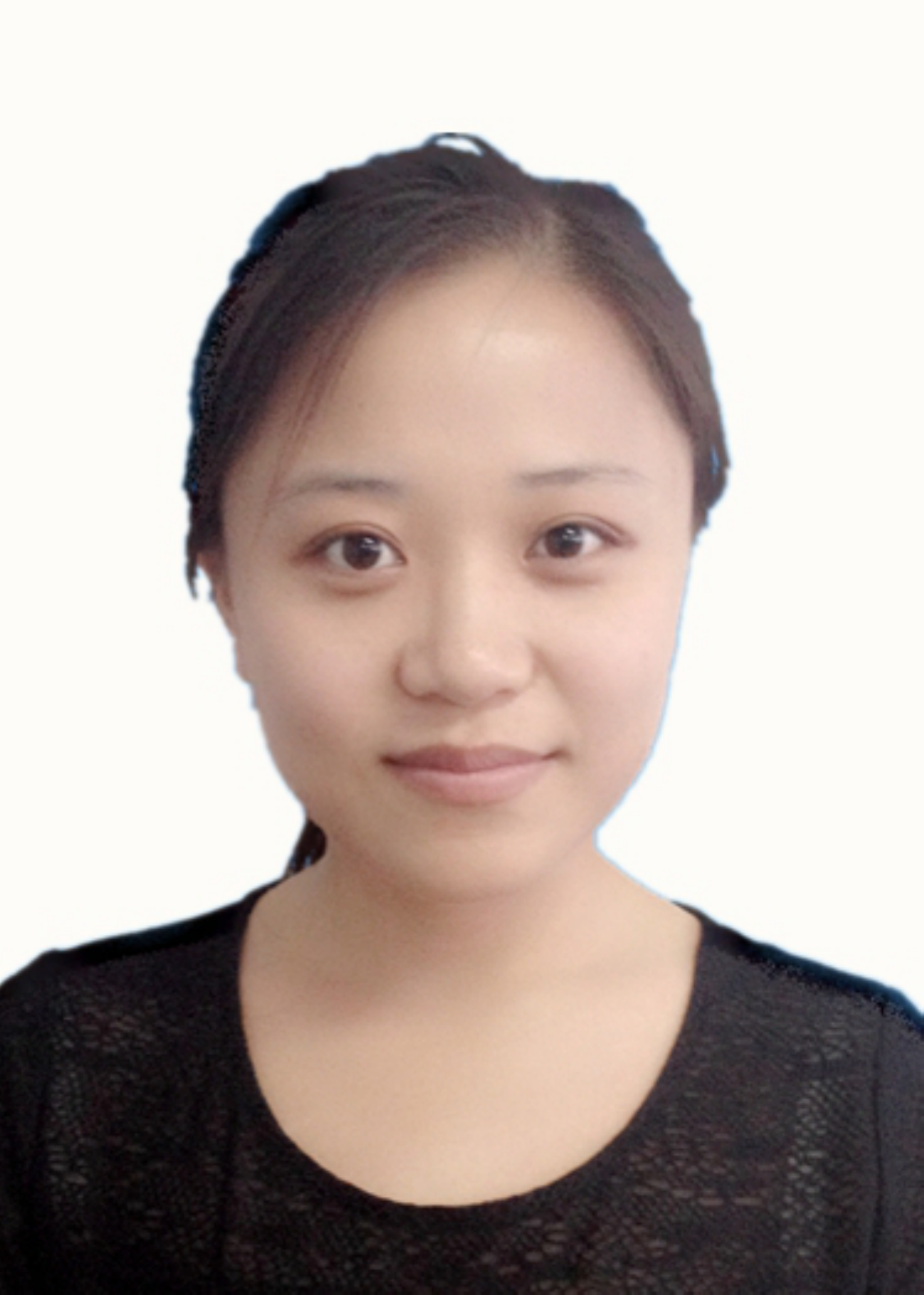}}]
{Zhen Lv} (M'21) received the B.Sc. and M.Sc. degrees in Information and Communication Engineering from Shandong University, China, in 2015 and 2018, respectively. She is currently a wireless channel engineer with the Purple Mountain Laboratories, Nanjing, China. Her research interests include 6G channel modeling and wireless big data.
\end{IEEEbiography}
\begin{IEEEbiography}[{\includegraphics[width=1.1in,height=1.3in,clip,keepaspectratio]{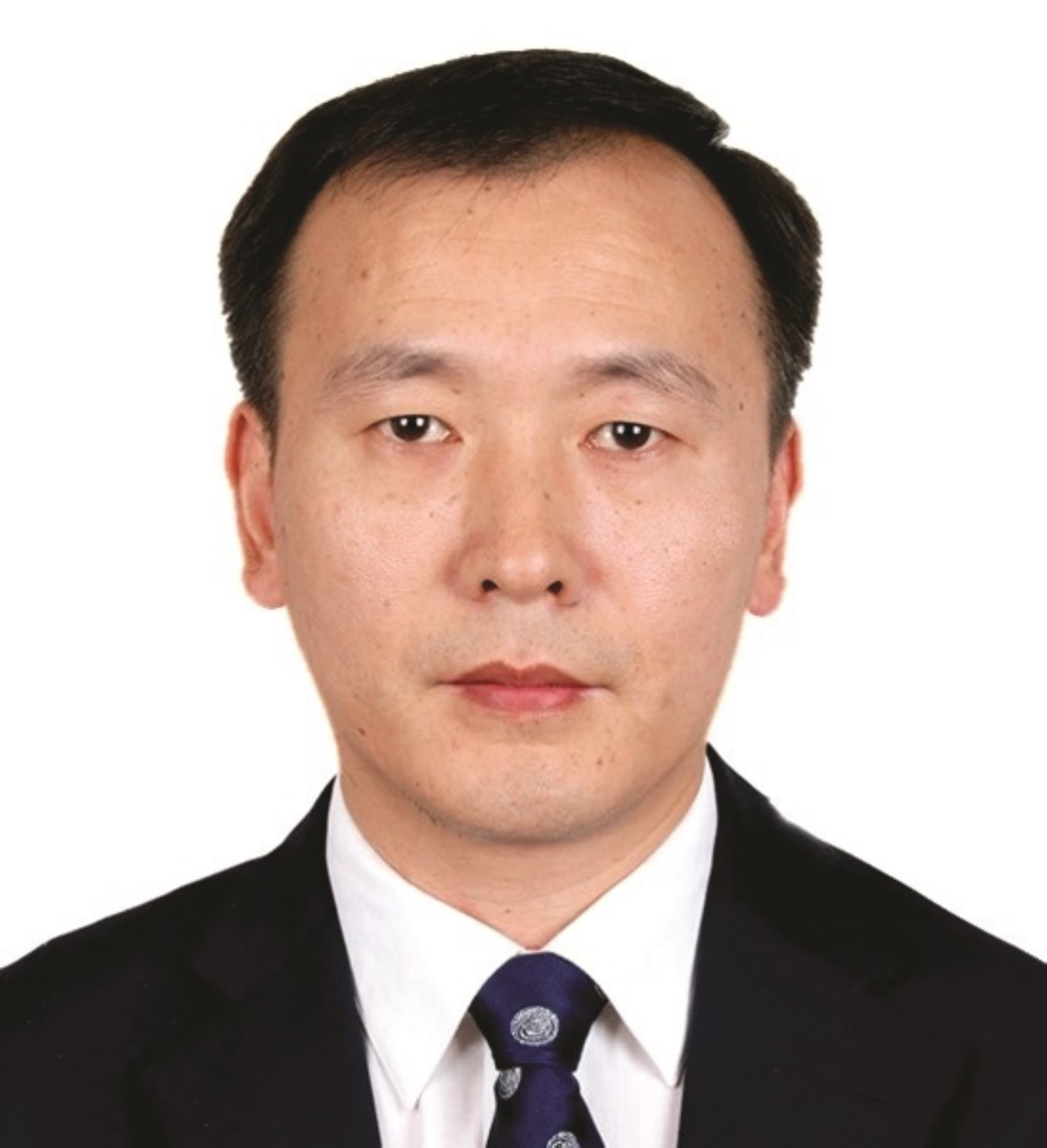}}]
{Xiqi Gao} (S'92-AM'96-M'02-SM'07-F'15) received the Ph.D. degree in electrical engineering from Southeast University, Nanjing, China, in 1997.

He joined the Department of Radio Engineering, Southeast University, in April 1992. Since May 2001, he has been a professor of information systems and communications. From September 1999 to August 2000, he was a visiting scholar at Massachusetts Institute of Technology, Cambridge, and Boston University, Boston, MA. From August 2007 to July 2008, he visited the Darmstadt University of Technology, Darmstadt, Germany, as a Humboldt scholar. His current research interests include broadband multicarrier communications, massive MIMO wireless communications, satellite communications, optical wireless communications, information theory and signal processing for wireless communications. From 2007 to 2012, he served as an Editor for the IEEE Transactions on Wireless Communications. From 2009 to 2013, he served as an Associate Editor for the IEEE Transactions on Signal Processing. From 2015 to 2017, he served as an Editor for the IEEE Transactions on Communications.

Dr. Gao received the Science and Technology Awards of the State Education Ministry of China in 1998, 2006 and 2009, the National Technological Invention Award of China in 2011, the Science and Technology Award of Jiangsu Province of China in 2014, and the 2011 IEEE Communications Society Stephen O. Rice Prize Paper Award in the field of communications theory.
\end{IEEEbiography}
\begin{IEEEbiography}[{\includegraphics[width=1.1in,height=1.3in,clip,keepaspectratio]{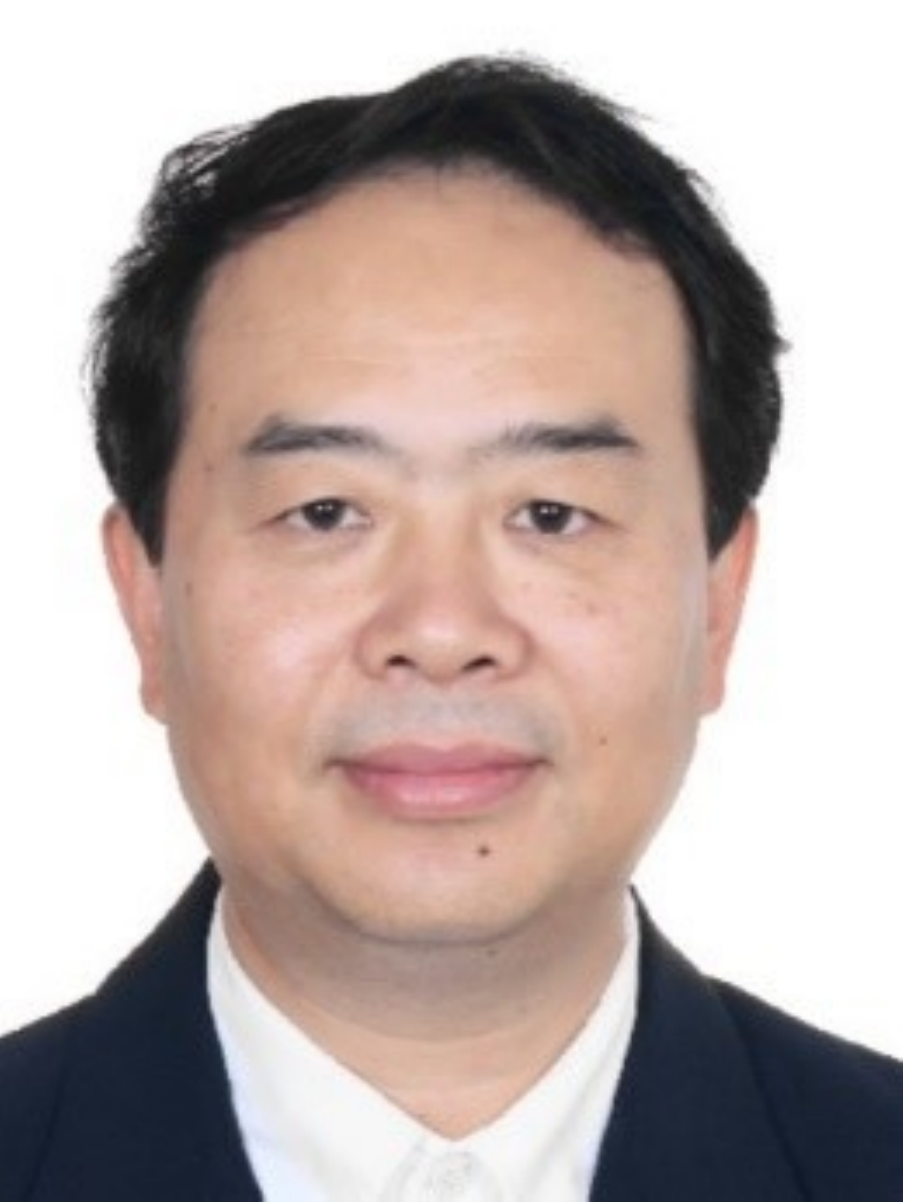}}]
{Xiaohu You} (SM'11-F'12)
has been working with National Mobile Communications Research Laboratory at Southeast University, where now he holds the rank of director and professor. He has contributed over 300 IEEE journal papers and 3 books in the areas of signal processing and wireless communications. From 1999 to 2002, he was the Principal Expert of the C3G Project. From 2001--2006, he was the Principal Expert of the China National 863 Beyond 3G FuTURE Project. Since 2013, he has been the Principal Investigator of China National 863 5G Project.

Professor You served as the general chairs of IEEE WCNC 2013, IEEE VTC 2016 Spring and IEEE ICC 2019. Now he is Secretary General of the FuTURE Forum, vice Chair of China IMT-2020 (5G) Promotion Group, vice Chair of China National Mega Project on New Generation Mobile Network. He was the recipient of the National 1st Class Invention Prize in 2011, and he was selected as IEEE Fellow in same year.
\end{IEEEbiography}
\begin{IEEEbiography}[{\includegraphics[width=1.1in,height=1.3in,clip,keepaspectratio]{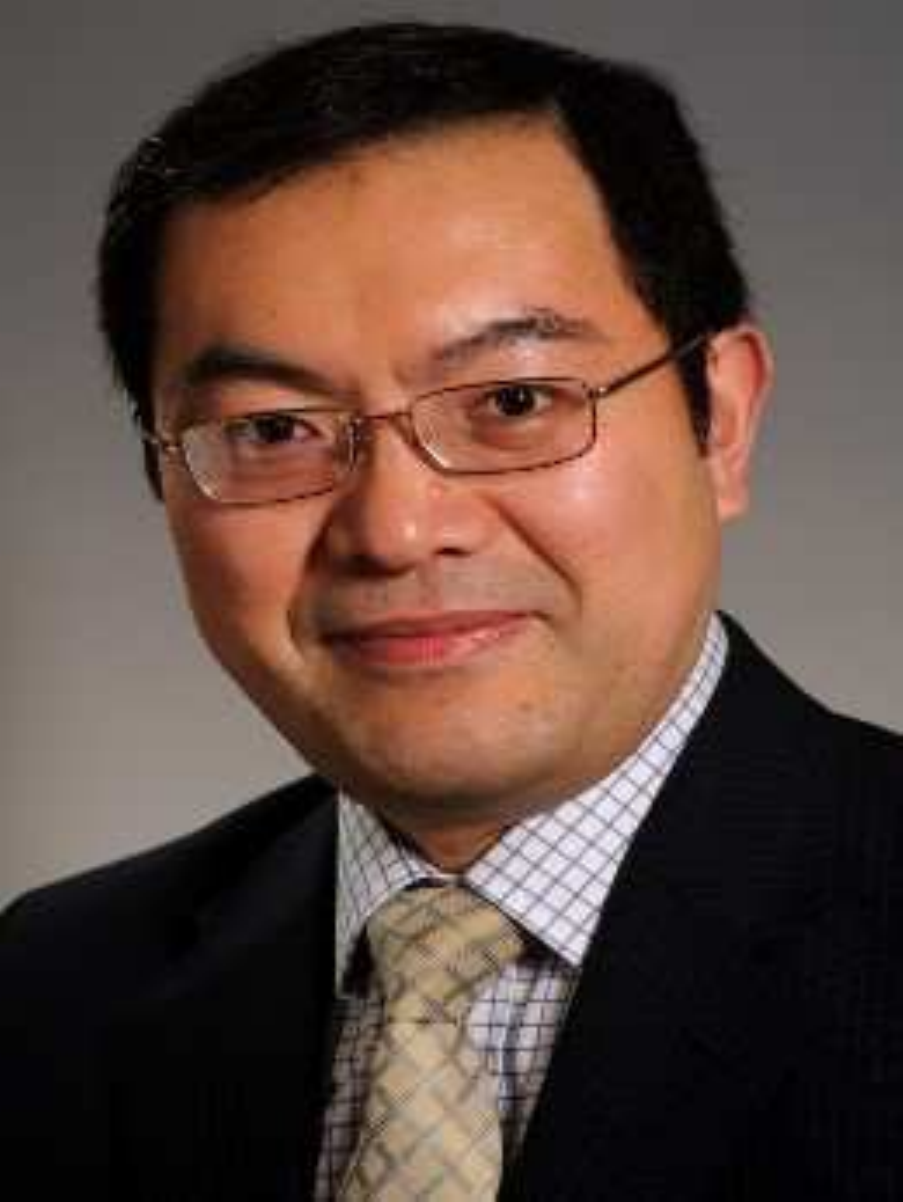}}]
{Yang Hao} (M'00-SM'06-F'13) received the Ph.D. degree in computational electromagnetics from the Centre for Communications Research, University of Bristol, Bristol, U.K., in 1998.

He was a Post-Doctoral Research Fellow with the School of Electronic, Electrical and Computer Engineering, University of Birmingham, Birmingham, U.K. He is currently a Professor of antennas and electromagnetics with the Antenna Engineering Group, Queen Mary University of London, London, U.K. He developed several fully integrated antenna solutions based on novel artificial materials to reduce mutual RF interference, weight, cost, and system complexity for security, aerospace, and healthcare, with leading U.K. industries, and novel and emergent gradient index materials to reduce mass, footprint, and profile of low-frequency and broadband antennas. He also co-developed the first stable active non-Foster metamaterial to enhance usability through small antenna size, high directivity, and tuneable operational frequency. He coined the term ��body-centric wireless communications,�� i.e., networking among wearable and implantable wireless sensors on the human body. He was the first to characterize and include the human body as a communication medium between on-body sensors using surface and creeping waves. He contributed to the industrial development of the first wireless sensors for healthcare monitoring. He has authored or coauthored more than 200 journal papers and has co-edited and coauthored the books Antennas and Radio Propagation for Body-Centric Wireless Communications (Artech House, 2006 and 2012) and FDTD Modeling of Metamaterials:
Theory and Applications (Artech House, 2008), respectively. His current research interests include computational electromagnetics, microwave metamaterials, graphene and nanomicrowaves, antennas and radio propagation for bodycentric wireless networks, active antennas for millimeter/submillimeter applications, and photonic integrated antennas.

Dr. Hao won 2015 IET AF Harvey Research Prize and is a co-recipient of BAE Chairman��s Silver Award in 2014 and the Royal Society Wolfson Research Merit Award. He is an elected Fellow of Royal Academy of Engineering, IET and IEEE.
\end{IEEEbiography}
\begin{IEEEbiography}[{\includegraphics[width=1.1in,height=1.3in,clip,keepaspectratio]{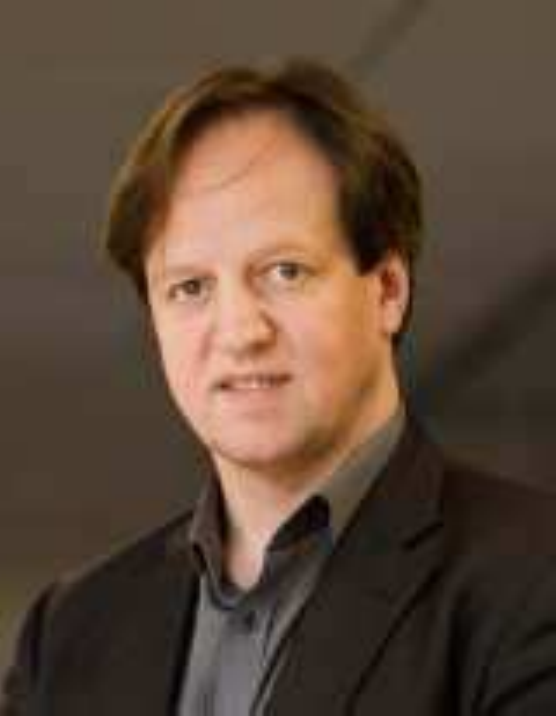}}]
{Harald Haas} (S'98-A'00-M'03-SM'16-F'18) received the Ph.D. degree from The University of Edinburgh in 2001. He is a Distinguished Professor of Mobile Communications at The University of Strathclyde/Glasgow, Visiting Professor at the University of Edinburgh and the Director of the LiFi Research and Development Centre. Prof. Haas set up and co-founded pureLiFi. He currently is the Chief Scientific Officer. He has co-authored more than 600 conference and journal papers. He
has been among the Clarivate/Web of Science highly cited researchers between 2017-2021. Haas' main research interests are in optical wireless communications and spatial modulation which he first introduced in 2006. In 2016, he received the Outstanding Achievement Award from the International Solid State Lighting Alliance. He was the recipient of IEEE Vehicular Society James Evans Avant Garde Award in 2019. In 2017, he received a Royal Society Wolfson Research Merit Award. He was the recipient of the Enginuity The Connect Places Innovation Award in 2021. He is a Fellow of the IEEE, the Royal Academy of Engineering (RAEng), the Royal Society of Edinburgh (RSE) as well as the Institution of Engineering and Technology (IET).
\end{IEEEbiography}
\end{document}